\documentclass[aps,pre,preprint,showpacs,amsmath,amsfonts,preprintnumbers]{revtex4}
\usepackage{epsfig}
\newcommand{\Op}[1]{{{\mathrm{\hat{#1}}}}}

\begin{document}
\title{The rise and fall of quantum and classical correlations in an open-system dynamics.}
\author{Michael Khasin and Ronnie Kosloff }
\affiliation{Fritz Haber Research Center for Molecular Dynamics, 
Hebrew University of Jerusalem, Jerusalem 91904, Israel}
\date{\today }

\begin{abstract} Interacting  quantum systems evolving from an uncorrelated  composite initial state generically develop  quantum correlations $-$ entanglement. As a consequence, a local description of interacting quantum system is impossible as a rule. A unitarily evolving (isolated) quantum system generically develops \textit{extensive} entanglement: the magnitude of the generated entanglement will increase without bounds with the effective Hilbert space dimension of the system. It is conceivable, that coupling of the interacting subsystems to {\textit{local} dephasing environments} will restrict the  generation of entanglement to such extent, that the evolving composite system may be considered as \textit{approximately disentangled}. 
This conjecture is addressed in the context of some common models of a bipartite  system with linear and nonlinear interactions and local coupling to dephasing environments. Analytical and numerical results obtained  imply that  the conjecture is generally false.
Open dynamics of the quantum correlations is compared to the corresponding evolution of the classical correlations and a qualitative difference is found. 
 
\end{abstract}

\maketitle

\section{Introduction}

The exploration of the nature and the extent of correlations generated by the many-body 
dynamics has both fundamental and practical applications. 
One of the fundamental issues in the investigation of many-body dynamics is finding  an optimal set of coordinates \cite{Moiseyev83, Jungwirth99}. This problem is solved in classical mechanics by introducing partition of a complex system into smaller subsystems, i.e., introducing of degrees of freedom. Description of the composite system is furnished by local descriptions of the subsystems.  
The adequacy of  a particular partition depends heavily on the nature 
and the extent of the correlations between the local degrees of freedom. 

The role played by correlations in classical and quantum mechanics is substantially different. This is due to the presence of the  quantum correlations, or entanglement, in a composite quantum state, having no analog in the classical world \cite{peres}. In contrast to classical correlations \cite{werner}, extensive  entanglement  makes the partition of a quantum system meaningless, since local measurements do not provide information on the state of an entangled system \cite{peres}.

The problem of the optimal partition is deeply connected to the
foundation of many-body dynamical simulations. The complete description of system composed of fully correlated subsystems
should grow exponentially with the number of subsystems involved.
 A possibility of representing a state of a complex system as a mixture of independently evolving uncorrelated states (trajectories) solves in principle the problem of many-body simulations, permitting to sample single trajectories for simulation and averaging the result subsequently   \cite{Miller02,Heller06}. This possibility is inherent in classical mechanics but is nongeneric in quantum case, due to the fact that typical interaction of quantum system generates entanglement.
If the growth of total (i.e. quantum and classical) correlations becomes restricted, the quantum dynamics can be 
efficiently simulated \cite{Meyer00,vidal1,vidal2,zwolak,vidal06}. Nonetheless, it is still an open question whether restrictions on quantum correlations alone are sufficient to provide for efficient simulations \cite{Jozsa}.

Addressing the problem of dynamical generation of correlations it is necessary to distinguish between the unitary evolution of an isolated system and the open evolution of a system coupled to  an environment. While a given unitary evolution can generate extensive entanglement, coupling the system to an environment is generally expected to restrict entanglement generation. This expectation originates in the general philosophy, seeing in environmental-induced decoherence \cite{Joos} the universal route of quantum-to-classical transition.  It is consistent with some established results on open-systems entanglement dynamics.

 Evolution of quantum correlations under the influence of environment was investigated both in the context of quantum to classical transition \cite{Joos} and in the context of quantum information processing \cite{Nielsen}. Most studies have been concerned with dynamics of entanglement between \textit{noninteracting} systems coupled to  a bath. It was found that coupling to common environment is able to entangle noninteracting systems \cite{Braun,Benatti}. On the other hand, coupling  to certain {local} environments leads to total disentanglement of the systems in finite time \cite{ diosi03,Halliwell04, Dodd04,Eberly03,Eberly06,FrancaSantos}.  The rates of disentanglement were calculated in bi- and multipartite systems of  noninteracting qubits \cite{Eberly02,Eberly03,Eberly04,Buchleitner04, Dur} and quidits \cite{Buchleitner, Mintert}, locally coupled to various environments. 
A number of studies addressed dynamical generation of correlations between \textit{interacting} subsystems in the presence of the environment. Production of entanglement between qubits, modeling a system of ions, coupled to environment through their center of mass motion  in ion-traps, was investigated in Ref.\cite{Mintert}. It was found that the coupling to environment diminishes the maximally achievable entanglement, with the corresponding entanglement loss increasing with the number of ions.
Ref. \cite{Lee} explored  dynamics of entanglement in quantum Heisenberg XY chain, immersed in a global purely dephasing bath. The robustness of  entanglement against the dephasing was related to the number of spin in the chain. Coupling of  interacting subsystems to local environments was considered in Refs.\cite{Plenio04} and \cite{Dodd04}.  Ref. \cite{Plenio04} investigated  the generation and transfer of entanglement in harmonic chains. The creation  of entanglement by suddenly switching on the interaction in the chain was found robust against the decoherence induced by coupling of the oscillators to local harmonic baths. The model of two harmonically coupled quantum Brownian particles was treated in Ref.\cite{Dodd04}.  
It was found that in the physically interesting range of parameters the interaction between the particles cannot prevent their eventual disentanglement, induced by coupling to local baths.

While the observed disentanglement of the noninteracting systems by coupling to local environments meets the common intuition about the quantum-to-classical transition, the picture of dynamics of entanglement in the presence of  interaction is not so clear. Searching for an efficient and a universal environment-induced mechanism of restricting the extent of the generated entanglement, it seems necessary to focus on the following aspects of the dynamics of correlations. 

First,  the scaling of the generated correlations with the effective Hilbert space dimension of the interacting subsystems must be  considered. This is in contrast to the context of the quantum information processing where the  object of interest is usually the scaling of entanglement with the number of degrees of freedom (qubits). The expectation is that the environment-induced restriction on the generation of entanglement becomes most significant in the range of the large quantum numbers of the system, which is commonly associated with the quantum-to-classical transition. In fact, extensive entanglement in the large Hilbert space dimension seems impossible without creating the "cat-state" superpositions, which are expected to be destroyed by the decoherence. 

 Second, the dynamics of correlations must be followed on the short, interaction time scales. It is possible that the long-time dynamics of an open composite system, approaching  equilibrium,  is disentangled, but the entanglement  generated on the interaction time scales is so large that the partition of the system has no meaning.
 
 Moreover, since a \textit{common} environment will generically entangle noninteracting systems,  coupling to \textit{local} environments seems necessary to provide for a generic route to a disentangled dynamics. 
 
  The present study  focuses on the investigation of environment-induced constraints on the dynamics of quantum and classical correlations  in the open  bipartite composite system. The system consists of two nonlinearly interacting harmonic oscillators, coupled to  local purely dephasing baths. 
  The distinction between the dephasing and pure dephasing has first appeared in the context of NMR \cite{Abragam}. Pure dephasing  corresponds to loss of coherence in the energy representation. The two prototypes of underlying stochastic processes leading to dephasing are the Gaussian and the Poissonian processes \cite{Gardiner}. R. Kubo  based his line-shape theory \cite{kubo62} on the Gaussian model. The Kubo's model is the cornerstone of the condensed-matter spectroscopy. Recently exceptions to the Gaussian paradigm have been found experimentally \cite{Yamaguchi, k183, Rice06} in ultrafast vibrational spectroscopy. The Poissonian model was shown to describe the dynamics adequately. Quantum Poissonian stochastic models have first appeared in the gas collision theory. They are also employed in the condense phase physics. For example, the Poissonian noise has been considered as a source of decoherence in  quantum dots \cite{Uskov,Kammerer,San-Jose}. Due to the fundamental and the experimental relevance of the Gaussian and the Poissonian stochastic processes they were chosen as the source of the dephasing in the present study.   
  
  The models aim to explore  dynamics of correlations in a composite system of  coupled multilevel subsystems at large effective Hilbert space dimension. Examples of such systems include multimode molecular vibrations \cite{Iachello}, linear and nonlinear quantum optics \cite{Perina} and cold trapped atomic ions \cite{Wineland98}.The primary goal is to locate a generic mechanism by which the decoherence keeps an interacting composite system "approximately disentangled" all along the evolution. Dynamics of quantum and classical correlations and their scaling with the effective Hilbert space dimension of the system are compared.

The measures of quantum and  total, i.e. quantum and classical,  correlations 
are defined in Section II. 
Section III examines the issue of the generation of {quantum} correlations (entanglement)
in the model problems.  
Section IV presents numerical results on dynamics of both quantum and classical correlations and Section V summarizes the conclusions.

\section{Measures of correlation}
The state of a bipartite system is uncorrelated if it can be described by the form 
\begin{eqnarray}
\Op \rho_{ab}=  \Op \rho_a \otimes \Op \rho_b.
\end{eqnarray}

A general correlated state  can be Schmidt-decomposed \cite{schmidt} (Cf.   Appendix A ) in the Hilbert-Schmidt space leading to:
\begin{eqnarray}
\Op \rho_{ab}=\sum_i^N  c_i \Op A_a^i \otimes \Op B_b^i , 
\label{separability}
\end{eqnarray}
where the sets $\{ \Op A\}$ and $\{ \Op B \}$ of operators  are orthonormal in the Hilbert-Schmidt spaces of systems $a$ and $b$.

The number of non vanishing coefficients $c_i$ in the Schmidt decomposition  of a vector in a abstract tensor-product Hilbert space is called the \textit{Schmidt rank} of the vector. To avoid confusion in the following presentation  the term \textit{HS-Schmidt rank} (or just \textit{HS rank} for brevity) is adopted for the Schmidt decomposition in the  Hilbert-Schmidt (HS) space of operators, while retaining the term  Schmidt rank for the Schmidt decomposition in the corresponding  Hilbert (pure) state space.
A HS rank is a natural measure of \textit{total correlations} present in a mixed state $\Op \rho$ (Cf. Appendix A).

A special subset of mixed states is the set of separable or classically correlated states \cite{werner}.  The state is separable if it can be cast into the following form
\begin{eqnarray}
\Op \rho_{ab}=\sum_i^N  p_i \Op \rho_a^i \otimes \Op \rho_b^i , 
\label{eq:clcor}
\end{eqnarray}
where $0\leq p_i \leq 1$ and $\sum_i^N p_i=1$  and $\Op \rho_a$ and $\Op \rho_b$ are density 
operators defined on the Hilbert spaces of the subsystems $a$ and $b$, respectively. Separable states are mixtures of uncorrelated states, which can be completely characterized by local measurements. Therefore, partition of a composite system into parts has a strong physical meaning. Such a partition is always possible
for classical probability density distribution of a bipartite system \cite{werner, Diosi07}. 
 
 Pure correlated states are always entangled. The measure of pure state 
entanglement can be defined by its \textit{Schmidt rank} \cite{virmani}.
Estimating the  measure of mixed-state entanglement is a difficult conceptual and computational problem \cite{virmani}. One can look for decomposition of $\Op \rho_{ab}$ into a mixture of pure states that are least entangled on average. The average entanglement corresponding to such decomposition is a possible  measure of the mixed state entanglements. Unfortunately,  such measures are  notoriously difficult to compute.  

An alternative computable measure of the bipartite mixed state entanglement  is  the \textit{negativity}  \cite{vidal} defined as follows:
\begin{eqnarray}
{\cal N}(\Op \rho)\equiv \frac{\left\|\Op \rho^  {T_a}\right\|-1}{2},
\end{eqnarray}
where $\left\|\Op X \right\|= \texttt{Tr}\sqrt{\Op X^{\dagger}\Op X}$ is the trace norm 
of an operator $\Op X$ and $T_a$ stands for the partial transposition with respect 
to the first subsystem.  The partial transposition $T_a$,
with respect to subsystem $a$ of a bipartite state $\Op \rho_{ab}$ 
expanded in a local orthonormal basis as  
$\Op \rho_{ab}=\sum \rho_{ij,kl}\left|i \right\rangle \left\langle j \right|  \otimes \left|k \right\rangle \left\langle l \right| $,
is defined as:
\begin{eqnarray}
\rho_{ab}^  {T_a}\equiv \sum \rho_{ij,kl}\left|j \right\rangle \left\langle i \right|  \otimes \left|k \right\rangle \left\langle l \right|.
\end{eqnarray}
The spectrum of the partially transposed density matrix is independent  of the choice of 
local basis or on the choice of the subsystem with respect to which the partial 
transposition is performed.
The negativity of the  state equals the absolute value of the sum of  
the negative eigenvalues 
of the partially transposed state. By the Peres-Horodecki criterion \cite{peres96, Horodeckii} 
the negativity vanishes in a separable state. On the other hand, vanishing of the negativity does not imply separability of the state in general \cite{Horodeckii}.  

Finite negativity is  necessary and sufficient condition for the presence of entanglement in particular type of mixed states, the so called Schmidt-correlated states \cite{Rains,Rains_error, Virmani_sacchi}. In this case the negativity can be related to the structure of the density operator, which  facilitates the evaluation of the entanglement. 

The  Schmidt-correlated states have the following form 
 \begin{eqnarray}
 \Op \rho=\sum_{mn }\rho_{mn}\left|\phi_m \right\rangle \left\langle \phi_n \right|\otimes \left| \chi_m\right\rangle \left\langle \chi_n\right|,
\label{eq:almsep1}
\end{eqnarray} 
where  $\Xi_1=\left\{\left|\phi_m \right\rangle\right\}_{m=1}^k$ and $\Xi_2=\left\{\left|\chi_m\right\rangle\right\}_{m=1}^k$ are local orthonormal bases. Eq.(\ref{eq:almsep1})  implies that $\Op \rho=\sum_i p_i \left|\psi_i\right\rangle\left\langle \psi_i\right|$, where  $\left|\psi_i\right\rangle=\sum_m c_m^i \left|\phi_m\right\rangle\otimes\left|\chi_m\right\rangle_2$ for every $i$, i.e. all pure states in the mixture share the same Schmidt bases (Cf.   Appendix A ) $\Xi_1$ and $\Xi_2$. It has been proved \cite{Dani} that for Schmidt-correlated states
\begin{eqnarray}
 {\cal N}(\Op \rho)=\sum_{m<n}|\rho_{mn}|,
\label{eq:bellman}
\end{eqnarray}
i.e., the negativity equals half the sum of absolute values of the off-diagonal elements of the density operator, written in a $\Xi_1\otimes\Xi_2$ local tensor product basis. It follows that the negativity of entangled Schmidt-correlated states is finite \cite{Dani}. 

The negativity can be related to  the structure of the density operator. Consider the density operator (\ref{eq:almsep1}) having the following quasi diagonal structure:
\begin{eqnarray}
 \Op \rho= \sum_{|m-n| \le \Delta }\rho_{mn}\left|\phi_m \right\rangle \left\langle \phi_n \right|\otimes \left| \chi_m\right\rangle \left\langle \chi_n\right|,
\label{eq:almsep2}
\end{eqnarray} 
with $\Delta \ll k$. The sum of the absolute values of the off-diagonal elements can be estimated as follows:
 \begin{eqnarray}
 \sum_{m\neq n}|\rho_{mn}|&=&  \sum_{m n}|\rho_{mn}|-1 = \sum_{m}\sum_{n=m-\Delta}^{n=m+\Delta}|\rho_{mn}|-1 <  \sum_{m}\sum_{n=m-\Delta}^{n=m+\Delta}\sqrt{\rho_{mm}\rho_{nn}} \nonumber \\ \\
 &\le &\sum_{m}\sum_{n=m-\Delta}^{n=m+\Delta}\frac{\rho_{mm}+\rho_{nn}}{2}=\frac{1}{2}\sum_{m}\sum_{n=m-\Delta}^{n=m+\Delta}\rho_{mm}+\frac{1}{2}\sum_{m}\sum_{n=m-\Delta}^{n=m+\Delta}\rho_{nn} < 2\Delta, \nonumber
\label{eq:estimation}
\end{eqnarray}
where the first inequality follows from the positivity of the density operator and the second is the inequality of geometric and arithmetic means. Therefore,
\begin{eqnarray}
{\cal N}(\Op \rho)<\Delta 
\label{eq:negbound}
\end{eqnarray}
in the state (\ref{eq:almsep2}). Since the negativity of maximally entangled state (corresponding to $|\rho_{mn}|=1/k$ in Eq.(\ref{eq:almsep1})) equals $(k-1)/2$, as follows from Eq.(\ref{eq:bellman}), the negativity of quasi diagonal density matrices is negligible compared to the maximally entangled state.  It should be noted that the form (\ref{eq:almsep2}) with $\Delta \ll k$ of the density matrix does not constrain the magnitude of the classical correlations present in the state. For example, a strictly diagonal matrix $\rho_{mn}=\delta_{mn}$ corresponds to a maximally (classically) correlated separable state.

Schmidt correlated states appear naturally in a composite bipartite dynamics admitting particular conservations laws \cite{Dani}. The models of open-system dynamics considered in the following sections belong to that class. As a consequence, the presence and extent of entanglement in evolving composite systems can be related to the structure of the density matrix, which can be inferred on the basis of relatively general scaling considerations.

\section{Density operator of a bipartite system under local pure dephasing.}
\subsection{General considerations}

The model of  open system dynamics  considered is described by:
\begin{eqnarray}
\frac{\partial}{\partial t}\Op \rho~~=~~({\cal L}_1+{\cal L}_2)\Op \rho + {\cal I} \Op \rho,
\label{general}
\end{eqnarray} 
where the generators of local nonunitary evolution are ${\cal L}_{j}=-i[\Op H_{j},\bullet]-\Gamma_j {\cal D}_{j}$, $j=1,2$ and  ${\cal I} =- i\gamma  [\Op H_{12},\bullet ]$ stands for the interaction superoperator.  The operators $\Op H_j$  are local system Hamiltonians, the operator $\Op H_{12}$ is the nonlocal (interaction) term in the composite system Hamiltonian and
${\cal D}_{j}$ denote local bath-dependent  superoperators. Coupling constants $\Gamma_{1,2}$ and $\gamma$ measure respectively the  
 strength of coupling to the local environments and  the strength of the interaction between the  subsystems.  

As a reference, the open evolution of noninteracting subsystems ( $\gamma=0$) is considered first. 
In this case a local dephasing evolution of each separate system takes place:
\begin{eqnarray}
\frac{\partial}{\partial t}\Op \rho~~=~~({\cal L}_1+{\cal L}_2)\Op \rho.
\label{generallocal}
\end{eqnarray}

In many models of open evolution \cite{Zurek81,lajos00,strunz, Joos} it is found that the evolving state undergoes decoherence, characterized by the decay of the off-diagonal elements of the density operator in particular basis of the \textit{robust states}. Ideal robust states retain their purity notwithstanding the interaction with environment.   Examples of such models include interaction with a purely dephasing environment, singling out energy states as the robust basis, quantum Brownian motion and damped harmonic oscillator at zero temperature $T=0$, which select the robust basis of coherent states. While the robust states basis is determined by  the type of the bath and  the system Hamiltonian, the time scales of the decoherence generally depend on the initial state as well. 

Let us assume that the local superoperators ${\cal L}_1$ and  ${\cal L}_2$ in Eq.(\ref{generallocal}) single out local robust states bases $\Xi_1$ and $\Xi_2$.  A composite noninteracting system evolving according to Eq.(\ref{generallocal}) from an arbitrary initial state is expected to decohere in the  tensor product  basis: $\Xi_1\otimes\Xi_2$. That means that an arbitrary initial state density matrix will eventually diagonalize in this basis. 
Switching on the interaction between subsystems causes a competition between entanglement generation
and decoherence induced by the local baths. 
For sufficiently weak interaction viewing the evolving density operator in the unperturbed tensor product basis $\Xi_1\otimes\Xi_2$ of local robust states  is a good starting point.  If the interaction perturbs only slightly the evolution of an off-diagonal matrix element, it will decay on an almost unperturbed decoherence time scale.

To proceed with a more quantitative argument the concept of the effective Hilbert space ${\cal H}_{eff}$
is helpful. Since the energy of the  evolving system is finite, the evolution can be effectively restricted to a Hilbert space with finite dimension. This Hilbert space is termed the effective Hilbert space of the system.
Let $\lambda$ be a spectral norm \cite{Horn} of the interaction superoperator ${\cal I}$ restricted to the effective Hilbert-Schmidt space (i.e., the space of linear operators on ${\cal H}_{eff}$) and $\Lambda$ be a spectral norm of the dissipator ${\cal D}={\cal D}_{1}+{\cal D}_{2}$ restricted to this space.
$\lambda$ and $\Lambda$ correspond to the shortest time scales of the evolution  generated by the ${\cal I}$ and ${\cal D}$, respectively. 
When $\lambda \ll \Lambda$, the interaction timescale is slow compared to the shortest decoherence 
time scale. 
As a consequence, the evolution of certain matrix elements is only slightly perturbed by the interaction. In that case the perturbed dynamics of the matrix element will follow essentially the course of the decoherence.  Therefore, a rough   distinction can be made  between the region of the density matrix dominated by the decoherence and the region  dominated by the interaction. The border between the two regions is defined by the condition
\begin{eqnarray}
 \tau_{ij}=O(\lambda^{-1}),
\label{eq:width}
\end{eqnarray}
where $\tau_{ij}$ is the unperturbed decoherence time scale of a matrix element $\rho_{ij}$, $i,j\in \Xi_1\otimes\Xi_2$.

In case that the decoherence-dominated regions of the density matrix are not populated initially, they will stay unpopulated in the course of the perturbed evolution.  This property will shape the structure of the evolving density matrix. If the states are  Schmidt-correlated states with local Schmidt-bases $\Xi_1$ and $\Xi_2$, being the local robust states bases, the relation can be established between the structure of the matrix and  the entanglement of the state as indicated in the previous section. Qualitatively, the larger is the decoherence-dominated region the smaller is the negativity of the state. 

The relative extent of the decoherence- and the interaction-dominated regions in a given dynamics generally depends on the initial state and, in particular , on the effective Hilbert-space dimension $k$ of the system. As a consequence,  different scenarios may be expected at asymptotically large $k$. The growing contribution of the interaction-dominated regions will generally imply extensive entanglement generation. On the other hand, if the relative size of the interaction-dominated regions becomes negligible at large $k$ the  entanglement generated by the open system dynamics  may be negligible or even asymptotically independent on $k$. This possibility  appeals to one who believes in the environmental-induced decoherence as a universal instrument of   quantum to classical transition.  An interesting question is the fate of the classical correlations in this scenario. While decoherence-dominated dynamics can turn extensively entangled initial state into extensively classically correlated state, it is not clear that decoherence-dominated dynamics can generate extensive classical correlations when quantum entanglement is negligible all along the evolution. Negligible total correlations seem nongeneric and do not correspond to  the intuitive picture of a "really interacting" system. Therefore, a scenario of negligible quantum and extensive classical correlations matches best to a generic mechanism of quantum to classical transition.

\subsection{Model calculations}

The model calculations are used to illustrate and verify the general considerations presented above.
The  evolution of a bipartite system is studied according to Eq.(\ref{general}) 
$ \frac{\partial}{\partial t}\Op \rho~~=~~({\cal L}_1+{\cal L}_2)\Op \rho + {\cal I} \Op \rho,$ where
(${\cal L}_{j}=-i[\Op H_{j},\bullet]-\Gamma_j {\cal D}_{j}$, $j=1,2$ and  ${\cal I} =- i\gamma  [\Op H_{12},\bullet ]$ )
 with two types of  dissipators, corresponding to the Gaussian \cite{gorini76,k216} and the Poissonian \cite{k183} purely dephasing models:
 \begin{eqnarray}
{\cal D}_j\Op \rho&=&\left\{\begin{array}{cc} \left[\Op H_j,\left[\Op H_j,\Op \rho \right]\right]  \ \ \ (Gaussian)  \\  \  e^{-i\phi \Op H_j}\Op \rho e^{i\phi \Op H_j}-\Op \rho \ \ (Poissonian) \end{array} \right.
\label{eq:bathtype}
\end{eqnarray}
These dissipators have the Lindblad form \cite{Lindblad76} of a generators of quantum dynamical semigroups.
The Gaussian and the Poissonian generators are the two examples explicitely mensioned in the seminal paper by G. Lindblad \cite{Lindblad76}.

The model Hamiltonian is a simplified version of a nonlinearly interacting multimode system.
The local Hamiltonians  $\Op H_j$, $j=1,2$ are chosen to be Hamiltonians of harmonic oscillators: $\Op H_j=\omega_j \Op a_j^{\dagger}\Op a_j$, where $\Op a_j^{\dagger}$ and $\Op a_j^{\dagger}$ are the creation and annihilation operators, respectively.
Two general types of interaction are considered. The first (${\cal I}_{A}$),  termed \textit{band-limited interaction}, is motivated by the stimulated Raman interaction between the translational modes of ions in cold traps \cite{Agarwal, Monroe98, Knight}(for example, in Ref. \cite{Monroe98} the effective interaction between the modes is reduced to the band-limited operator $\exp[\pm i \pi(\Op a^{\dagger}_x\Op a_y+\Op a_x\Op a^{\dagger}_y)]$). 
The second type of interaction (${\cal I}_{B}$) is motivated by weakly nonlinear interacting modes emerging  in molecular vibrations \cite {Iachello} and in nonlinear optics. The typical  example in the nonlinear optics is the second harmonic generation modeled by the interaction Hamiltonian of the form $\Op H=\hbar g(\Op a^2 \Op b^{\dagger} +\Op a^{\dagger 2} \Op b)$ \cite{Agarwal, Perina}. In addition, the dynamics of the cold ion traps can be operated in the regime, where the effective interaction is well approximated by a weakly nonlinear coupling \cite{Agarwal, Monroe98, Knight}.

The two types of interaction are generated by: 
\begin{eqnarray}
{\cal I}&=&\left\{\begin{array}{cc} - i\gamma  [\Op A_{1}^{\dagger}\Op A_{2}+\Op A_{2}^{\dagger}\Op A_{1},\bullet ] \ (\equiv {\cal I}_A) \\  \  - i\gamma  [(\Op a_{1}^{\dagger})^s(\Op a_{2})^r+(\Op a_{2}^{\dagger})^s(\Op a_{1})^r ,\bullet ]\ \ \ s=1,2,..;\ r=1,2,..,  \ (\equiv {\cal I}_B) \end{array}  \right. 
\label{eq:interaction}
\end{eqnarray}
where $ \Op A_j$ is defined by its matrix elements in local energy basis: $(\Op A_j)_{mn}=\delta_{m, n-1}$. The structure of $\Op A_j$ assures that  ${\cal I}_{A}$ is band-limited with the spectral norm $\lambda =O(\gamma)$. 

The important property of the dynamics Eq.(\ref{general}), with local dephasing Eq.(\ref{eq:bathtype}) and interaction Eq.(\ref{eq:interaction}) is conservation of a particular additive operator in each case. The first type of interaction, ${\cal I}_A$, preserves the number operator 
$\Op N=\Op a_{1}^{\dagger}\Op a_{1}+\Op a_{2}^{\dagger}\Op a_{2}$: ${\cal I}_A^{\dagger}(\Op N)=0$, which is also preserved by the local generators: $({\cal L}_1^{\dagger}+{\cal L}_2^{\dagger})(\Op N)=0$. The second type of interaction, ${\cal I}_B$, preserves the generalized number operator $\Op N_{rs}\equiv r\Op a_{1}^{\dagger}\Op a_{1}+s \Op a_{2}^{\dagger}\Op a_{2}$: ${\cal I}_B^{\dagger}(\Op N_{rs})=0$, preserved by the local generators, as well: $({\cal L}_1^{\dagger}+{\cal L}_2^{\dagger})(\Op N_{rs})=0$.

Assume a pure uncorrelated initial state $\left|\psi(0)\right\rangle = \left|k 0\right\rangle$ (written in the local energies basis).  The state $\left|\psi(0)\right\rangle$ is an eigenstate of  $\Op N$ with the eigenvalue $k$.  As a consequence, the first type of the interaction, ${\cal I}_A$, will drive the  initial state   into a mixture of eigenstates
of $\Op N$ corresponding to the  eigenvalue $k$: 
$\Op \rho(t)=\sum_{mn}c_{mn}\left|m  \ l_m\right\rangle\left\langle n  \ l_n\right|$ with $l_m=k-m$. 
Thus, $k$ determines the effective Hilbert space dimension of the system in this case: $\texttt{dim}({\cal H}_{eff})=k$. 
Since $\left|\psi(0)\right\rangle$ is also an eigenstate of  $\Op N_{rs}$ with the eigenvalue $rk$, the  second type of interaction, ${\cal I}_B$,  will take it  into a mixture of eigenstates
of  $\Op N_{rs}$ corresponding to the same eigenvalue $rk$: 
$\Op \rho(t)=\sum_{mn}c_{mn}\left|m  \ l_m\right\rangle\left\langle n  \ l_n\right|$ with  $l_m=\frac{r}{s}(k-m)$. 
The number of initial excitations $k$ of the first oscillator determines the effective Hilbert space 
dimension (in the strong sense) of the system in this case: $\texttt{dim}({\cal H}_{eff})=k/s$. To summarize:
\begin{eqnarray}
\Op \rho(t)=\sum_{mn}c_{mn}\left|m  \ l_m\right\rangle\left\langle n  \ l_n\right|\left\{\begin{array}{cc} l_m = k-m  \ \ ({\cal I}_A ) \\  \  l_m = \frac{r}{s}(k-m)  \ ({\cal I}_B) \end{array}  \right. 
\label{eq:solution}
\end{eqnarray}
In both cases, the resulting mixed state is a Schmidt-correlated state with a time-independent Schmidt bases. This property permits evaluation of the negativity of $\Op \rho$ in each case from its structure, as indicated in Section II.

 \subsubsection{Gaussian vs. Poisson pure dephasing bath}
The difference between the two types  (\ref{eq:bathtype}) of environments can be understood 
from comparing the local evolutions of a single oscillator coupled to the bath of each type:
\begin{eqnarray}
\frac{\partial}{\partial t}\Op \rho~~=~~-i[\Op H,\Op \rho]-\Gamma {\cal D}\Op \rho.
\label{local}
\end{eqnarray} 
In the Gaussian  case, the Eq.(\ref{local}) in the energy representation becomes:
\begin{eqnarray}
\dot{\rho}_{nm}=-i\omega_{mn}\rho_{nm}-\Gamma \omega_{mn}^2 \rho_{nm},
\end{eqnarray} 
with $\omega_{mn}\equiv\omega_m-\omega_n$, leading to the solution $\rho_{nm}(t)=\rho_{nm}(0)e^{-i\omega_{mn}t-\Gamma \omega_{mn}^2t}$. 
Thus the effect of the purely dephasing Gaussian bath is the "diagonalization" 
of the density matrix in the energy basis  (the robust states basis for this model)
on the time scale that varies for different matrix elements $\rho_{nm}$ 
and increases with the distance $|m-n|$ of the element from the diagonal. 
The shortest decoherence time scale corresponds to the largest distance from the diagonal and  decreases with the growing effective 
Hilbert space dimension of the system.

In the Poissonian case, the Eq.(\ref{local}) in the energy representation becomes:
\begin{eqnarray}
\dot{\rho}_{nm}=-i\omega_{mn}\rho_{nm}+\Gamma (e^{-i\omega_{mn}\phi}-1)\rho_{nm},
\end{eqnarray} 
leading to the solution $\rho_{nm}(t)=\rho_{nm}(0)e^{-i\omega_{mn}t+\Gamma (e^{-i\omega_{mn}\phi}-1)t}$.  Apparently, the robust states basis is once again the energy basis, but the  decoherence rates of the matrix elements are limited by $\left|\Gamma (e^{-i\omega_{mn}\phi}-1)\right|=2\Gamma$, independently of the initial state.
  
This difference in properties of the Gaussian and the Poissonian environments will result in different dynamics of the correlations in the composite bipartite  system dynamics Eq.(\ref{general}). 
\begin{figure}[t]
\epsfig{file=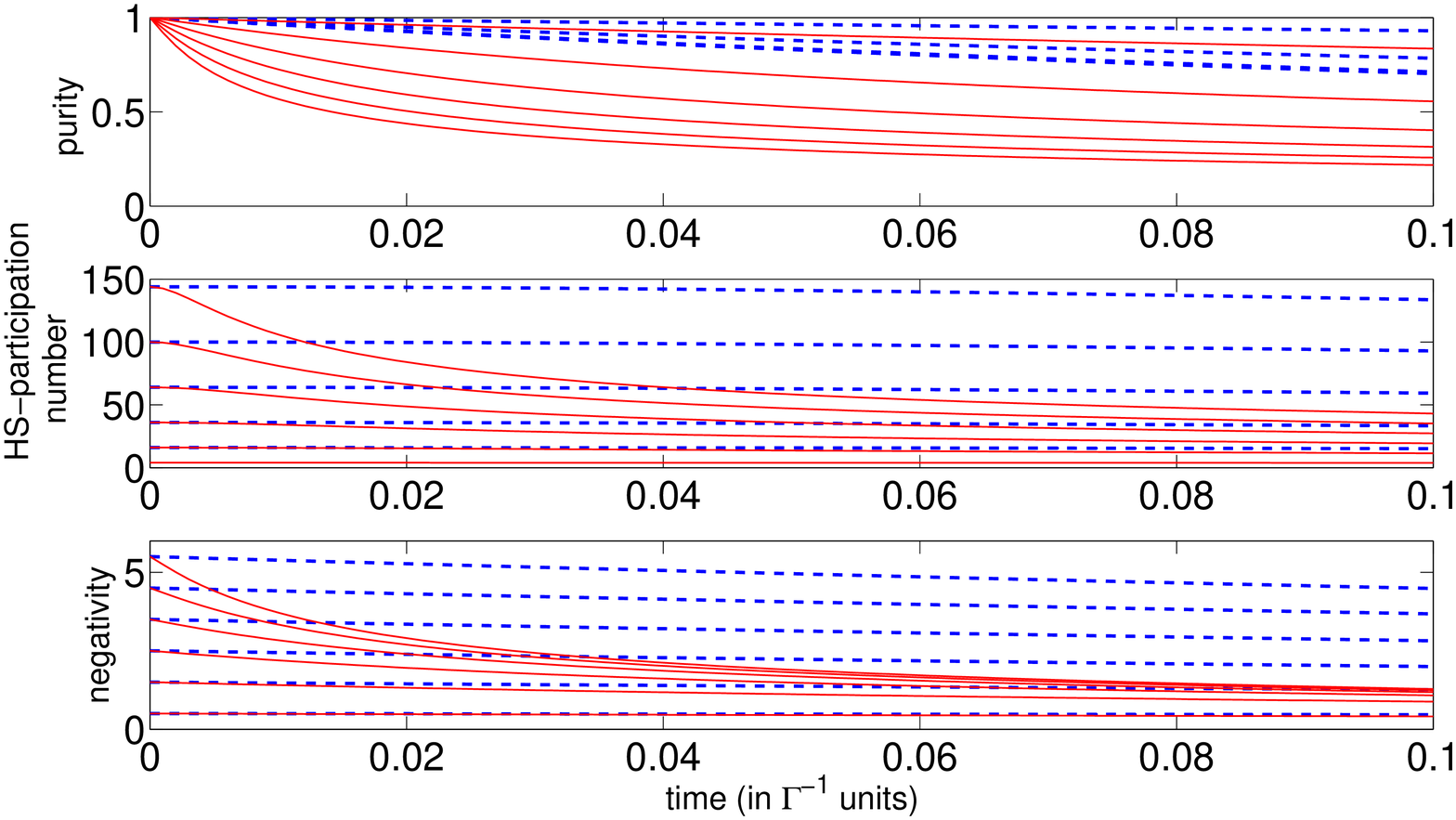, width=18.0cm, clip=} 
\caption{(Color online) Purity, HS-participation number and  negativity of the density operator of two  harmonic oscillators, evolving under local purely dephasing Gaussian (solid lines) and Poissonian (dashed lines) environments (Eq.(\ref{eq:bathtype})). The initial state is a pure maximally correlated  state $\left|\psi\right\rangle=\frac{1}{\sqrt{k+1}}\Sigma_{n=0}^k \left|n\right\rangle\left|k-n\right\rangle$ for $k=2,4,...,12$. The coupling parameter to the bath $\Gamma=1$ in both cases. The frequencies of the oscillators $\omega_1=\omega_2=1$. While the decay rates in the Gaussian case  depend on initial state and increase with the effective Hilbert space dimension $k$, in the Poissonian case the rates are practically independent of $k$.}
\label{fig:bathdriven}
\end{figure}

\subsubsection{Local dephasing driven dynamics}  
To gain insight on  the effect of dephasing on the correlations 
the simplest bath-driven dynamics is studied first, in which 
the composite system Hamiltonian vanishes altogether, meaning that  no entanglement is generated during the evolution. 
The corresponding Eq.(\ref{general}) transforms into 
 \begin{eqnarray}
\frac{\partial}{\partial t}\Op \rho~~&=&~~-\sum_{j=1,2}\Gamma_j \left[\Op H_j,\left[\Op H_j,\Op \rho \right]\right] \ \ \ \ \ \ \ \ \ (Gaussian \ dephasing), \label{localbathg} \\
\frac{\partial}{\partial t}\Op \rho~~&=&~~-\sum_{j=1,2}\Gamma_j \left(e^{-i\phi \Op H_j}\Op \rho e^{i\phi \Op H_j}-\Op \rho \right)\ (Poissonian \ dephasing).
\label{localbathp}
\end{eqnarray}

Since the dynamics preserves local energies the effective Hilbert space dimension of the evolving 
system is determined by the energy range of the initial state. The  initial state 
of the form 
 $\left|\psi\right\rangle=\frac{1}{\sqrt{k+1}}\sum_{n=0}^k\left|n\right\rangle\left|k-n\right\rangle$ ( $\left|n\right\rangle$ 
is a local energy eigenstate)  is a maximally entangled state. It corresponds to the effective Hilbert space spanned by the states $\left\{ \left|n\right\rangle\left|k-n\right\rangle \right\}_{n=0}^k$.

The solution to Eq. (\ref{localbathg}), the Gaussian case,  is found:
 \begin{eqnarray}
\Op \rho(t)&=& \frac{1}{k+1}\sum_{mn} e^{-(\Gamma_1\omega_{1,mn}^2+\Gamma_2\omega_{2,mn}^2)t}\left|n\right\rangle\left|k-n\right\rangle \left\langle m\right|\left\langle k-m\right| \label{solutiong}
\end{eqnarray}
when the solution to  Eq.(\ref{localbathp}),  the Poissonian case, becomes:
 \begin{eqnarray} 
\Op \rho(t)&=& \frac{1}{k+1}\sum_{mn} e^{-(\Gamma_1 [1-e^{-i\omega_{1,mn}\phi}]+\Gamma_2 [1-e^{-i\omega_{1,mn}\phi}])t}\left|n\right\rangle\left|k-n\right\rangle \left\langle m\right|\left\langle k-m\right|.
\label{solutionp}
\end{eqnarray}
 Decoherence rates in the Gaussian case (\ref{solutiong}) are $\tau_{mn}^{-1}=\Gamma_1\omega_{1,mn}^2+\Gamma_2\omega_{2,mn}^2\le \Lambda_g \equiv \texttt{max}_{m,n \le k}\left\{\Gamma_1\omega_{1,mn}^2+\Gamma_2\omega_{2,mn}^2\right\}$ and generally increase without bounds with the effective Hilbert space dimension $k$. For example, taking $\Op H_j=\omega_j \Op a^{\dagger}_j \Op a_j$, $\tau_{mn}^{-1}=(\Gamma_1\omega_1^2+\Gamma_2\omega_2^2)(m-n)^2$ is obtained, with maximal rate $\Lambda_g=\tau_{0k}^{-1}=(\Gamma_1\omega_1^2+\Gamma_2\omega_2^2)k^2$.  In the Poissonian case (\ref{solutionp}) the decoherence rates are bounded: $\tau_{mn}^{-1}=\texttt{Re}\left\{\Gamma_1 [1-e^{-i\omega_{1,mn}\phi}]+\Gamma_2 [1-e^{-i\omega_{1,mn}\phi}]\right\}\le \Lambda_p\equiv 2(\Gamma_1+\Gamma_2)$.  

Note, that both solutions (\ref{solutiong}) and (\ref{solutionp}) are Schmidt-correlated states. Therefore,
the corresponding negativities can be  calculated from Eq. (\ref{eq:bellman}) :
 \begin{eqnarray}
{\cal N}(\Op \rho(t))~~&=&\frac{1}{k+1}\sum_{m<n} e^{-(\Gamma_1\omega_{1,mn}^2+\Gamma_2\omega_{2,mn}^2)t}  \ \ \ \ \ \ \ \ \ \ \ \ (Gaussian \ dephasing), \label{negativityg} \\
{\cal N}(\Op \rho(t))~~&=&\frac{1}{k+1}\sum_{m<n} e^{-(\Gamma_1 [1-e^{-i\omega_{1,mn}\phi}]+\Gamma_2 [1-e^{-i\omega_{1,mn}\phi}])t}\ (Poissonian \ dephasing).
\label{negativityp}
\end{eqnarray}

 From Eqs.(\ref{negativityg}) and (\ref{negativityp}) it follows that both types of the purely dephasing dynamics ( Eqs.(\ref{localbathg}) and (\ref{localbathp})) lead eventually to
a complete decay of the quantum correlations (note that since the evolving state is Schmidt-correlated, its negativity vanishes if and only if the state is disentangled \cite{Dani}). But the dependence of the time scales of the decay on the effective Hilbert space dimension $k$ is different in the two cases. In the Poissonian case  the rate of the negativity (\ref{negativityp}) decay is bounded by $\Lambda_p=2(\Gamma_1+\Gamma_2)$, independent of  $k$, while in the Gaussian case (\ref{negativityg}), the bound is $\Lambda_g=\texttt{max}_{m,n<k}\left\{\Gamma_1\omega_{1,mn}^2+\Gamma_2\omega_{2,mn}^2\right\}$, which generally grows with $k$. 

The total correlations (and, as a consequence, the classical correlations) follow a different course of evolution. The HS-rank (and HS-participation number) of initial state is $k^2$ (see Appendix A for calculation of HS-rank of a pure state). The stationary solution corresponding to both Eqs.(\ref{solutiong}) and (\ref{solutionp}) is $\Op \rho_{st}= \frac{1}{k+1}\sum_{m}\left|m\right\rangle\left|k-m\right\rangle \left\langle m\right|\left\langle k-m\right|$ with HS-rank (and HS-participation number) equal to $k$. Therefore, although the total correlations decay  in both models, the stationary solution contains extensive classical correlations, i.e. the correlations that grow without bounds with the effective Hilbert space dimension $k$.

Fig. (\ref{fig:bathdriven}) displays the negativity, HS-participation number and purity of the composite state 
evolving under Gaussian (\ref{localbathg}) and Poissonian (\ref{localbathp}) dephasing dynamics, corresponding to $\Op H_i=\omega \Op a_i^{\dagger}\Op a_i$, $\Gamma_1=\Gamma_2$ and  initial state of the form $\left|\psi\right\rangle=\frac{1}{\sqrt{k+1}}\sum_{n=0}^k\left|n\right\rangle\left|k-n\right\rangle$. The effective 
Hilbert space dimension is varied: $k=4,...,12$. As anticipated from the difference of the two 
types of environments, the decay rates in the Gaussian case depend on the initial state 
and increases with the effective Hilbert space dimension, while in the Poissonian case 
the rates are effectively independent of the initial state.  

\subsubsection{Full dynamics} 
At this point the interaction between the oscillators are introduced and the full dynamics according to Eq.(\ref{general}) with $\gamma \neq 0$ is followed.
We shall consider a pure uncorrelated initial state of the composite system: $\left|\psi(0)\right\rangle=\left|k 0\right\rangle$, i.e. the state corresponding to the excitation of the $k$'th level of the first oscillator and the ground state of the second. In that case, as shown above, each type of the interaction (\ref{eq:interaction}) and the dephasing (\ref{eq:bathtype}) considered admits a particular additive conserved quantity (a generalized number operator), which defines the effective Hilbert space of the composite system for each $k$ and is responsible for the remarkable property of the evolving state: the density operator is a Schmidt-correlated state in a time-independent Schmidt bases: $
 \Op \rho(t)=\sum_{mn }\rho_{mn}(t)\left|m \right\rangle \left\langle n \right|\otimes \left| l_m\right\rangle \left\langle l_n\right|$ (Eq.(\ref{eq:solution})).  The Schmidt bases  $\Xi_1=\left\{ \left|m\right\rangle \right\}$ and $\Xi_2=\left\{ \left|l_m\right\rangle \right\}$ are the  robust (local energies) bases of the corresponding local open systems (\ref{local}), with the correspondence $m \leftrightarrow l_m$, determined by the particular conservation law, depending on the type of interaction. 
This property allows us  to relate the structure of the evolving density operator to its negativity,  as indicated in Section II.
The relevant structure of the evolving density operator is determined by the relative size of the decoherence- and the interaction-dominated regions of the corresponding density matrix. This structure is investigated  for each types of interaction and dephasing and for different effective Hilbert space dimensions of the system. 
 
 The overview in the preceding section of the dynamics driven solely by the local dephasing reveals important difference between the two types of local environment with respect to the anticipated structure of the evolving density matrix. In the Poissonian case the decoherence rates are of the order of the system-bath coupling: $\tau_{mn}^{-1}\le 2(\Gamma_1+\Gamma_2)$, as shown above. Therefore, evolution of the  matrix elements is dominated either by the decoherence or by the interaction depending on the relative strength of the coupling constants and independently of the effective Hilbert space dimension.  
In models with weak system-bath coupling, $\Gamma_{1,2} \ll \lambda$, the structure of the evolving density operator will only slightly be effected by the coupling 
to the Poissonian bath on  the interaction time-scale  $\gamma^{-1} \ll \tau_{mn}$. 
As a consequence, the quantum correlations will develop almost unperturbed on the interaction time scale.
 
 A different dynamical pattern is anticipated in the case of the Gaussian purely dephasing bath. The decoherence rates in this case are
 \begin{eqnarray}
\tau_{mn}^{-1}=\Gamma_1\omega_1^2(m-n)^2+\Gamma_2\omega_2^2(l_m-l_n)^2\left\{\begin{array}{cc} l_m = k-m  \ \ ({\cal I}_A ) \\  \  l_m = \frac{r}{s}(k-m)  \ ({\cal I}_B) \end{array}  \right. 
\label{eq:rateg}
\end{eqnarray}
where ${\cal I}_A $ and ${\cal I}_B$ indicate the type of interaction: ${\cal I}_A\equiv-i\gamma[\Op A_{1}^{\dagger}\Op A_{2}+\Op A_{2}^{\dagger}\Op A_{1},\bullet]$, with $(\Op A_j)_{mn}=\delta_{m, n-1}$, and  ${\cal I}_B\equiv-i\gamma[(\Op a_{1}^{\dagger})^s(\Op a_{2})^r+(\Op a_{2}^{\dagger})^s(\Op a_{1})^r,\bullet]$, 
with $(\Op a_j)_{mn}=\sqrt{m}\delta_{m, n-1}$, (see Eq.(\ref{eq:interaction})). In each case, the decoherence rate increases with the "distance" $|m-n|$ from the diagonal. As a consequence, the evolving density operator obtains a quasi-diagonal structure in the local energies basis, with the width $\Delta$ of the interaction-dominated region  about the diagonal depending on the type of interaction.

 Let us assume for simplicity that $\Gamma_1 \omega_1^2=\Gamma_2 \omega_2^2=\Gamma$. In that case a matrix element $\rho_{mn}$  decoheres on the time scale $\tau_{mn}=[2\Gamma (m-n)^2]^{-1}$. The spectral norm of  ${\cal I}_A$  is $\lambda=O(\gamma)$. Therefore,
 the width about the diagonal of the evolving density matrix can be estimated from Eq.(\ref{eq:width}) as 
 \begin{eqnarray}
\Delta=O(\sqrt{\gamma/\Gamma}),
\label{eq:scaleA}
\end{eqnarray}
where  $\Gamma \ll \gamma$ is assumed. The spectral norm of ${\cal I}_B$ is $\lambda=O(k^{\frac{r+s}{2}})$, 
where $k/s$ is the effective Hilbert space dimension of the system (see Eq.\ref{eq:solution}). 
As a consequence, from Eq.(\ref{eq:width}) $\Delta$ becomes:
 \begin{eqnarray}
\Delta=O(\sqrt{\gamma/\Gamma}k^{\frac{r+s}{4}}).
\label{eq:scaleB}
\end{eqnarray} 
In the band limited interaction case (${\cal I}_A$) the quasidiagonal structure of the density 
operator emerges (Eq.(\ref{eq:scaleA})), while in the case of the nonlinear interaction 
(${\cal I}_B$) a quasidiagonal structure is expected only if the nonlinearilty is weak: 
$s+r<4$ (Eq.(\ref{eq:scaleB})).

Perturbation theory  supports the scaling considerations. For a normalized eigenoperator $\Op O_l$ of the local evolution generator ${\cal L}_1^{\dagger}+{\cal L}_2^{\dagger}$: $({\cal L}_1^{\dagger}+{\cal L}_2^{\dagger})\Op O_l=\lambda_l \Op O_l$. The interaction $\cal I$ perturbs the evolution. The action of the perturbed generator on $\Op O_l$ gives: $({\cal L}_1^{\dagger}+{\cal L}_2^{\dagger}+{\cal I}^{\dagger})\Op O_l=\lambda_l (\Op O_l +\Op \delta_l)$. If the trace norm of  $\Op \delta_l$ is  small compared to unity: $\left\|\Op \delta_l\right\|_1\ll \left\|\Op O_l \right\|_1=1$, the evolution of $\Op O_l$ is only slightly perturbed on the time scale of $(\lambda_l)^{-1}$. Therefore, if $\texttt{Re}[\lambda_l]< 0$, the perturbed  $\Op O_l$ will  decay on a time scale of $\left|\texttt{Re}[\lambda_l]\right|$ to the leading order in $\left\|\Op \delta_l\right\|_1$.
 To each  density matrix element $\rho_{mn}$ in the nonperturbed tensor-product basis of the local energy states (the robust states bases) there corresponds the normalized operator $\Op O_{mn}=\left|m l_m\right\rangle\left\langle n l_n\right|$ such that $\left\langle \Op O _{mn}\right\rangle=\texttt{Tr}\left\{ \Op \rho  \Op O_{mn} \right\}=\rho_{mn}$. Defining $\Op \delta_{mn}$ by $({\cal L}_1^{\dagger}+{\cal L}_2^{\dagger}+{\cal I}^{\dagger})\Op O_{mn}=\lambda_{mn} (\Op O_{mn} +\Op \delta_{mn})$ with $\lambda_{mn}=i(\omega_1(m-n)+\omega_2(l_m-l_n))-\Gamma [(m-n)^2+(l_m-l_n)^2]$ we  obtain for the trace norm of $\Op \delta_{mn}$ corresponding to the first type of interaction ${\cal I}_A$:
 \begin{eqnarray}
\left\|\Op \delta_{mn}\right\|_1&=&O\left(\frac{\gamma}{\Gamma }\sqrt{\frac{1}{[\frac{\omega_1}{\Gamma}(m-n)]^2+[\frac{\omega_2}{\Gamma}(l_m-l_n)]^2+[(m-n)^2+(l_m-l_n)^2]^2}}\right) \nonumber \\
&<& O\left(\frac{\gamma}{\Gamma }\frac{1}{(m-n)^2+(l_m-l_n)^2}\right)
\end{eqnarray} 
and  for the trace norm of $\Op \delta_{mn}$ corresponding to the second type of interaction ${\cal I}_B$:
\begin{eqnarray}
\left\|\Op \delta_{mn}\right\|_1&=&O\left(\frac{\gamma}{\Gamma }\sqrt{\frac{n^rm^s+l_m^s l_n^r}{[\frac{\omega_1}{\Gamma}(m-n)]^2+[\frac{\omega_2}{\Gamma}(l_m-l_n)]^2+[(m-n)^2+(l_m-l_n)^2]^2}}\right) \nonumber \\
&<& O\left(\frac{\gamma}{\Gamma }\frac{\sqrt{l_m^rm^s+n^s l_n^r}}{(m-n)^2+(l_m-l_n)^2}\right).
\end{eqnarray} 

The width $\Delta$  of the interaction-dominated region is estimated by solving
 \begin{eqnarray}
 \frac{\gamma}{\Gamma }\frac{1}{(m-n)^2+(l_m-l_n)^2}=1
  \label{eq:scaleA1}
\end{eqnarray} 
for the interaction generated by ${\cal I}_A$ and 
 \begin{eqnarray}
 \frac{\gamma}{\Gamma }\frac{\sqrt{l_m^rm^s+n^s l_n^r}}{(m-n)^2+(l_m-l_n)^2}=1
 \label{eq:scaleB1}
\end{eqnarray} 
for the interaction generated by ${\cal I}_B$.
Using Eq.(\ref{eq:rateg}), Eq.(\ref{eq:scaleA1}) is simplified to:
\begin{eqnarray}
 \frac{\gamma}{2\Gamma }\frac{1}{(m-n)^2}=1,
 \label{eq:scaleA2}
\end{eqnarray} 
from which $\Delta=2|m-n|=\sqrt{2\gamma/\Gamma }$ is found, in compliance with the estimate (\ref{eq:scaleA}), and  Eq.(\ref{eq:scaleB1}) is simplified to:
 \begin{eqnarray}
 \frac{\gamma}{2\Gamma }\left(\frac{r}{s}\right)^{r/2}\frac{\sqrt{[(k-m)^rm^s+n^s (k-n)^r]}}{2(m-n)^2}=1.
 \label{eq:scaleB2}
\end{eqnarray}
In this case, the width about the diagonal $\Delta=2|m-n|$ will depend on $m$. 
The upper bound on $\Delta$ was  calculated from Eq.(\ref{eq:scaleB2}) in two cases. 
First, for the linear coupling $r=s=1$ gives $\Delta < 2^{3/4}\sqrt{\frac{\gamma k}{\Gamma}}$. 
Second, for the nonlinear coupling $r=1$, $s=2$ gives  $\Delta < \sqrt{\frac{\gamma}{\Gamma}}k^{3/4}$. 
Both results  comply with the estimation Eq.(\ref{eq:scaleB}).

\begin{figure}[t]
\epsfig{file=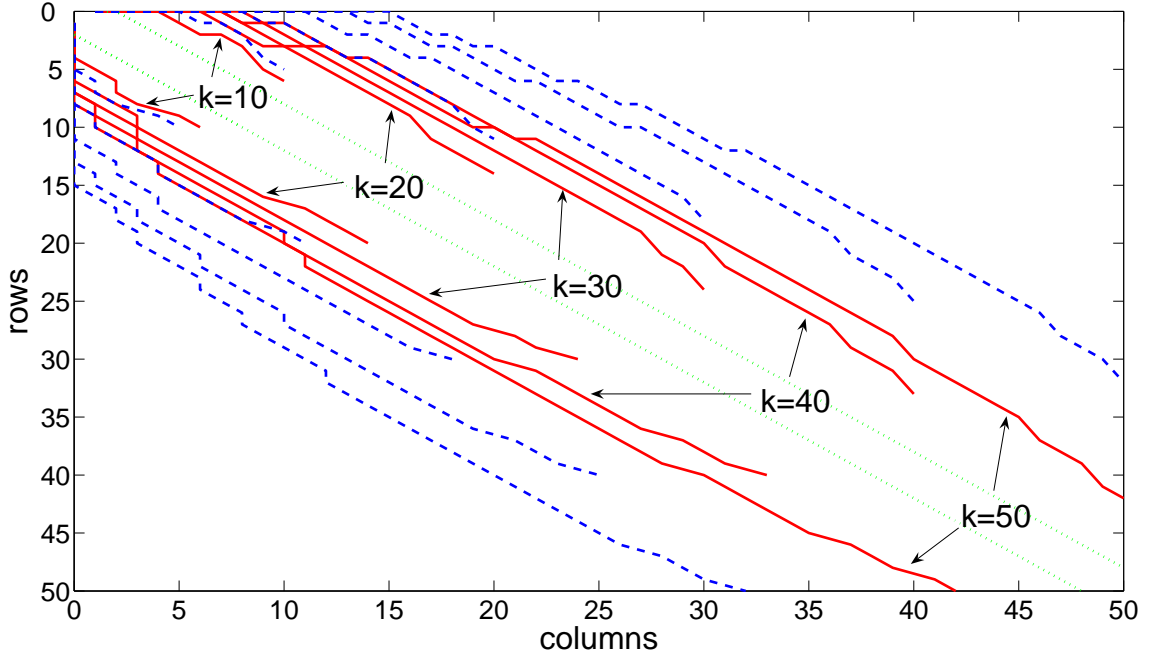, width=18.0cm, clip=} 
\caption{(Color online) The density operator of the state, evolving according to Eqs. (\ref{general}) for various interactions (\ref{eq:interaction}) and Gaussian type local baths (\ref{eq:bathtype}) is represented in the product of local energies bases (Schmidt bases). Boundaries are indicated, in each case, separating the outer (off-diagonal) regions, dominated by the decoherence, from the inner (near diagonal) interaction-dominated regions.
  The interactions correspond to the band-limited case (case $A$, Eq.(\ref{eq:interaction}), dotted lines), linear coupling $r=s=1$ (case $B$, Eq.(\ref{eq:interaction}), solid lines) and the nonlinear coupling $r=1$, $s=2$ (case $B$, Eq.(\ref{eq:interaction}), dashed lines). The density matrices in the band-limited case are of the form $\Op \rho=\Sigma_{mn}c_{mn}\left|m  \ k-m\right\rangle\left\langle n  \ k-n\right|$ and in the linear and nonlinear cases $\Op \rho=\Sigma_{mn}c_{mn}\left|m  \ (r/s)(k-m)\right\rangle\left\langle n  \ (r/s)(k-n)\right|$. The effective Hilbert space dimension corresponds to $k=10,20,30,40,50$ in each case. See explanations in the text.}
\label{fig:width}
\end{figure}
Figure (\ref{fig:width}) displays regions of the density matrix, dominated by the interaction, vs. regions, dominated by the decoherence, with the boundary between the regions determined by  Eqs.(\ref{eq:scaleA2}) 
and (\ref{eq:scaleB2}) for  $k=10,20,40,50$ and $\gamma/\Gamma=3$. 
The figure represents the composite system density matrices 
$\Op \rho=\sum_{mn}c_{mn}\left|m  \ k-m\right\rangle\left\langle n  \ k-n\right|$ 
and $\Op \rho=\sum_{mn}c_{mn}\left|m  \ \frac{r}{s}(k-m)\right\rangle\left\langle n  \ \frac{r}{s}(k-n)\right|$, 
corresponding to Eqs.(\ref{eq:scaleA2}) and (\ref{eq:scaleB2}), 
with $m$ indexing the columns and $n$ indexing the rows. 
The contours of  Eqs. (\ref{eq:scaleB2}) are plotted for the linear coupling ($r=s=1$) and the  
nonlinear coupling ($r=1$, $s=2$). The quasidiagonal structure of the density operator 
is apparent.  Both in the case of linear and nonlinear 
coupling between the oscillators, the width grows with the effective Hilbert space dimension.
This is in contrast to the case of 
the band-limited interaction (${\cal I}_A$), where the width about the diagonal 
does not depend on the effective Hilbert space dimension $k$.

To conclude, in contrast to the Poissonian type dephasing, in the Gaussian case the interaction-dominated regions are located about the diagonal of the density operator represented in the local energies basis. Since the initial state $\left|\psi(0)\right\rangle=\left|k 0\right\rangle$ corresponds to the density operator with an unpopulated decoherence-dominated region, this region will remain unpopulated all along the evolution. As a consequence, the evolving density operator will stay in the quasidiagonal form. According to Eq.(\ref{eq:negbound}) the value of negativity is bounded by $\Delta$ in each case: ${\cal N}(\Op \rho)< \Delta$.
Asymptotically, i.e. as $k\gg 1$ for the band-limited interaction, $\sqrt{k}\gg 1$ for the linear interaction and 
$\sqrt[4]{k}\gg 1$ for the nonlinear case, the width about the diagonal becomes negligible compared to $k$. In this case,  the generated entanglement
is negligible compared to the maximal entanglement compatible with the effective Hilbert space dimension.

In the following section results of numerical calculations of the evolution of negativity, illustrating the foregoing discussion, are presented. The evolution of negativity is compared in each case with dynamics of the total (i.e., quantum and classical) correlations, as measured by the effective HS-rank and HS-participation number  of the evolving density operator.

\begin{figure}[t]
\epsfig{file=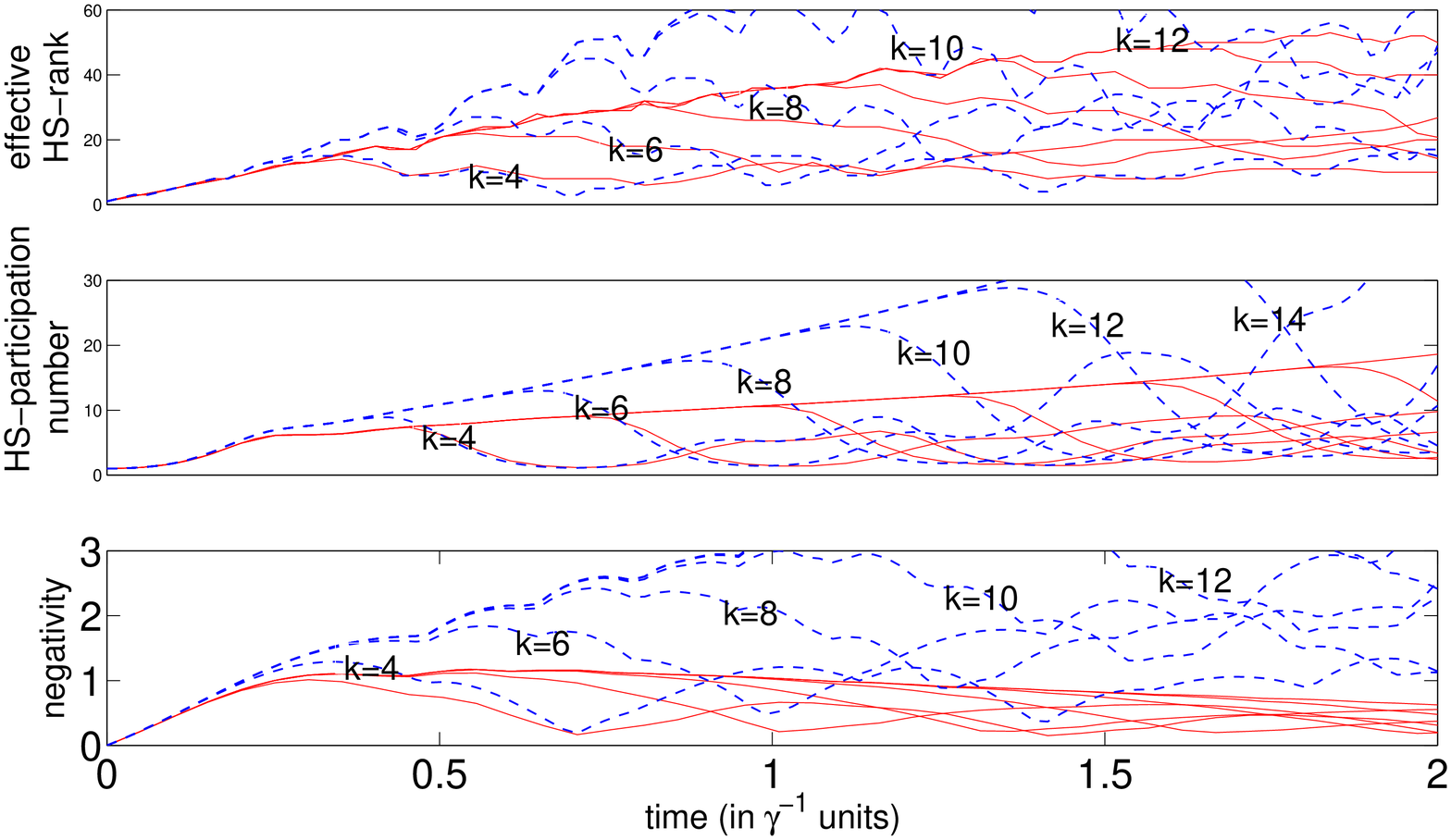, width=18.0cm, clip=} 
\caption{(Color online)The negativity, the effective HS-rank  and the HS-participation number of the  evolving density 
operator: cases $AG$ (solid lines) and $A$ (dashed lines). In both cases $\omega_1=\omega_2=\omega$, $\Gamma_1=\Gamma_2=\Gamma$, with $\Gamma \omega^2=(1/3)\gamma=(1/15)\omega$ in case $AG$ and $\Gamma=0$ in case $A$. Initial state $\left|\psi\right\rangle=\left| k \  0\right\rangle$, with $k=4,6,...,14$.}
\label{fig:AG}
\end{figure} 
 
\begin{figure}[t]
\epsfig{file=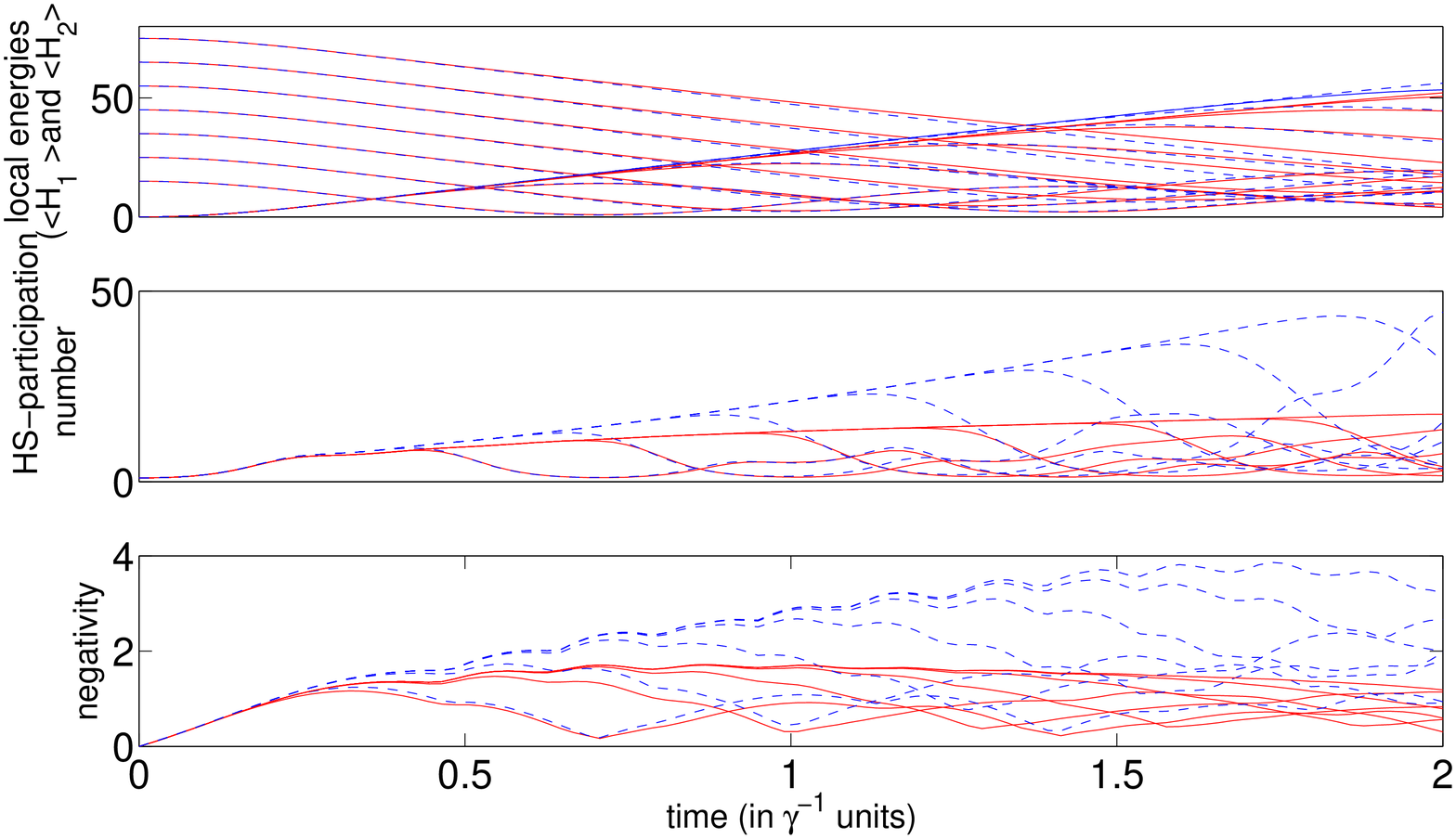, width=18.0cm, clip=} 
\caption{(Color online) The negativity, the local energies  and the HS-participation number of the  density 
operator: cases $AG$ and $AP$. Parameters: $\omega_1=\omega_2=\omega$, $\Gamma_1=\Gamma_2=\Gamma$ in both cases, $\Gamma \omega^2=0.125\gamma=0.025\omega$ in $AG$ (solid lines) and $\Gamma=(1/15)\gamma=(1/75)\omega$, $\phi=2 \pi/7$ in $AP$ (dashed lines). Initial state $\left|\psi\right\rangle=\left| k \  0\right\rangle$, with $k=4,6,...,14$.}
\label{fig:AGP}
\end{figure} 

\begin{figure}[t]
\epsfig{file=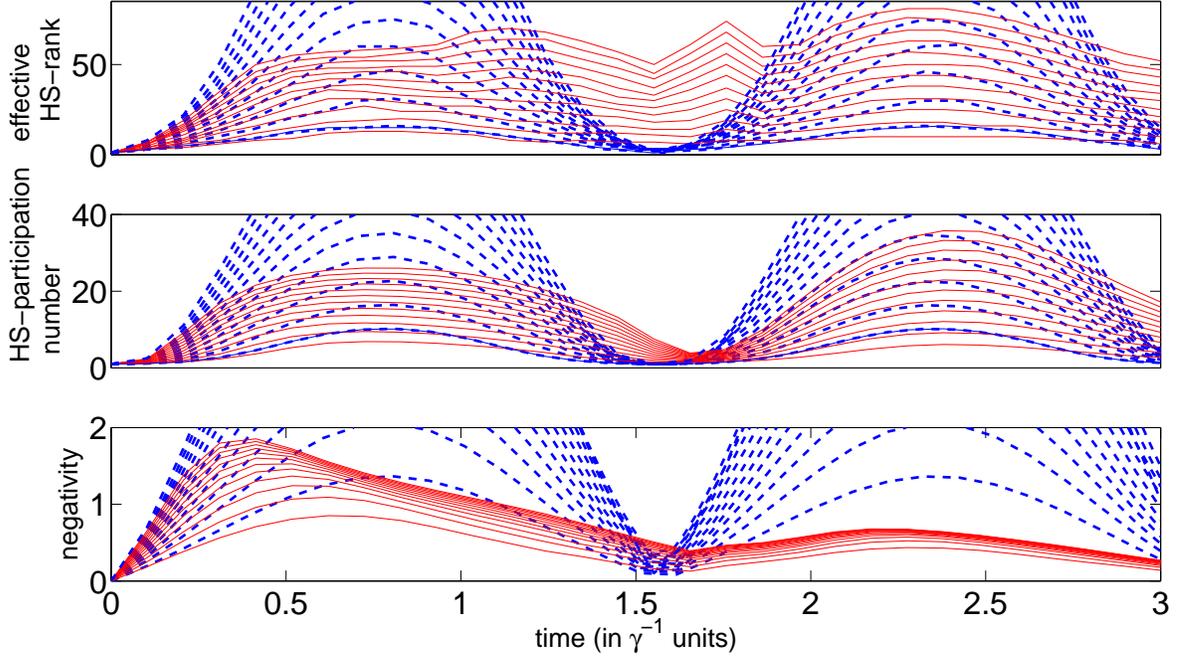, width=18.0cm, clip=} 
\caption{(Color online) The negativity, the effective HS-rank  and the HS-participation number of the  evolving density 
operator: cases $BG1$ (solid lines) and $B1$ (dashed lines). In both cases: $\omega_1=\omega_2=\omega$, in $BG1$ case:  $\Gamma_1=\Gamma_2=\Gamma$, $\Gamma \omega^2=(1/3)\gamma=(1/15)\omega$ and in $B1$ case: $\Gamma_1=\Gamma_2=0$. Initial state $\left|\psi\right\rangle=\left|k \ 0 \right\rangle$ for $k=4,6,...,24$. }
\label{fig:BG1}
\end{figure}

\begin{figure}[t]
\epsfig{file=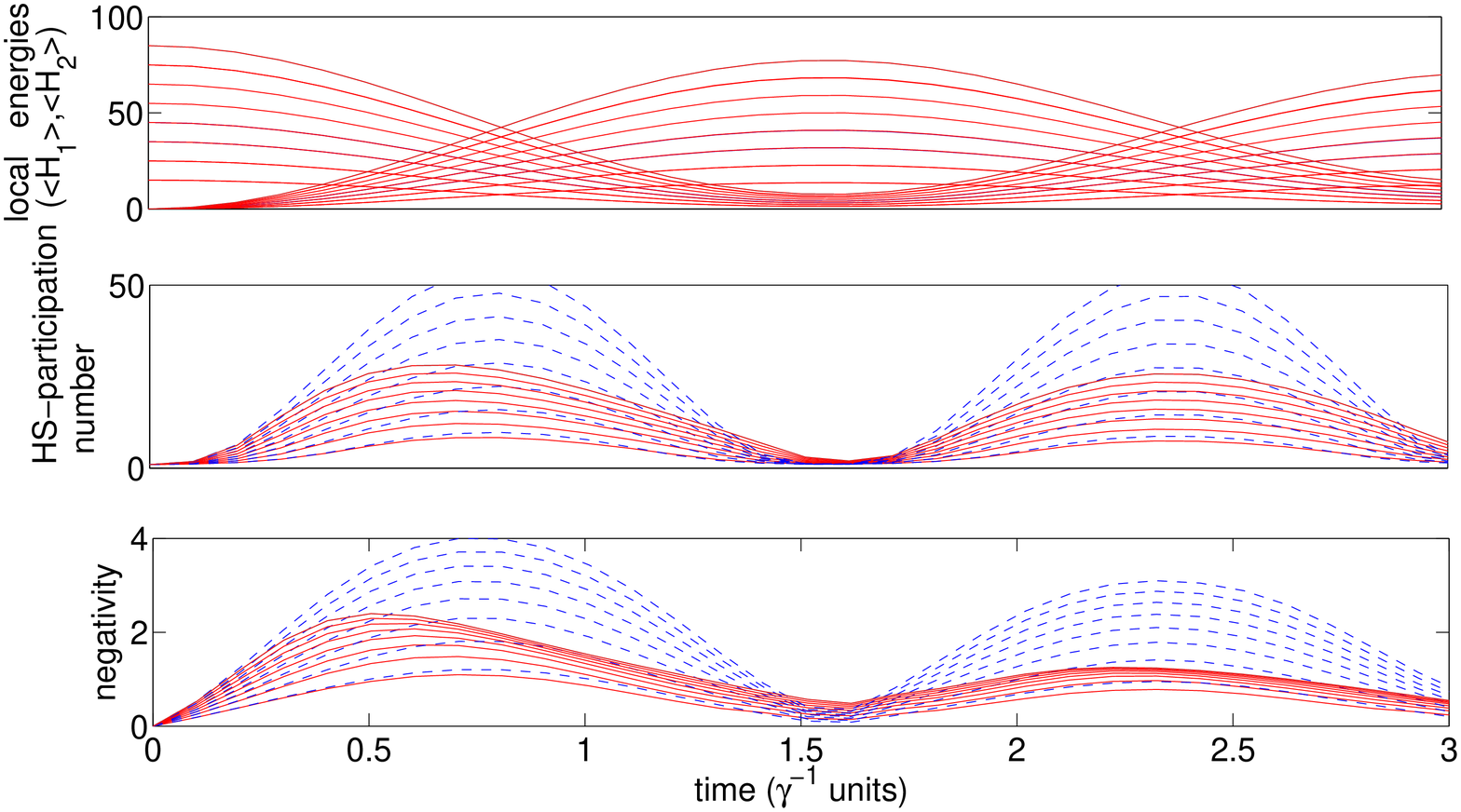, width=18.0cm, clip=} 
\caption{(Color online) The negativity, the local energies  and the HS-participation number of the  density 
operator: cases $BG1$ and $BP1$. Parameters: $\omega_1=\omega_2=\omega$, $\Gamma_1=\Gamma_2=\Gamma$ in both cases, $\Gamma \omega^2=(1/16)\gamma=(1/80)\omega$ in $BG1$ (solid lines) and $\Gamma=(1/10)\gamma=(1/50)\omega$, $\phi=2 \pi/7$ in $BP1$ (dashed lines). Initial state $\left|\psi\right\rangle=\left| k \  0\right\rangle$, with $k=4,6,...,18$. }
\label{fig:BGP1}
\end{figure}

\begin{figure}[t]
\epsfig{file=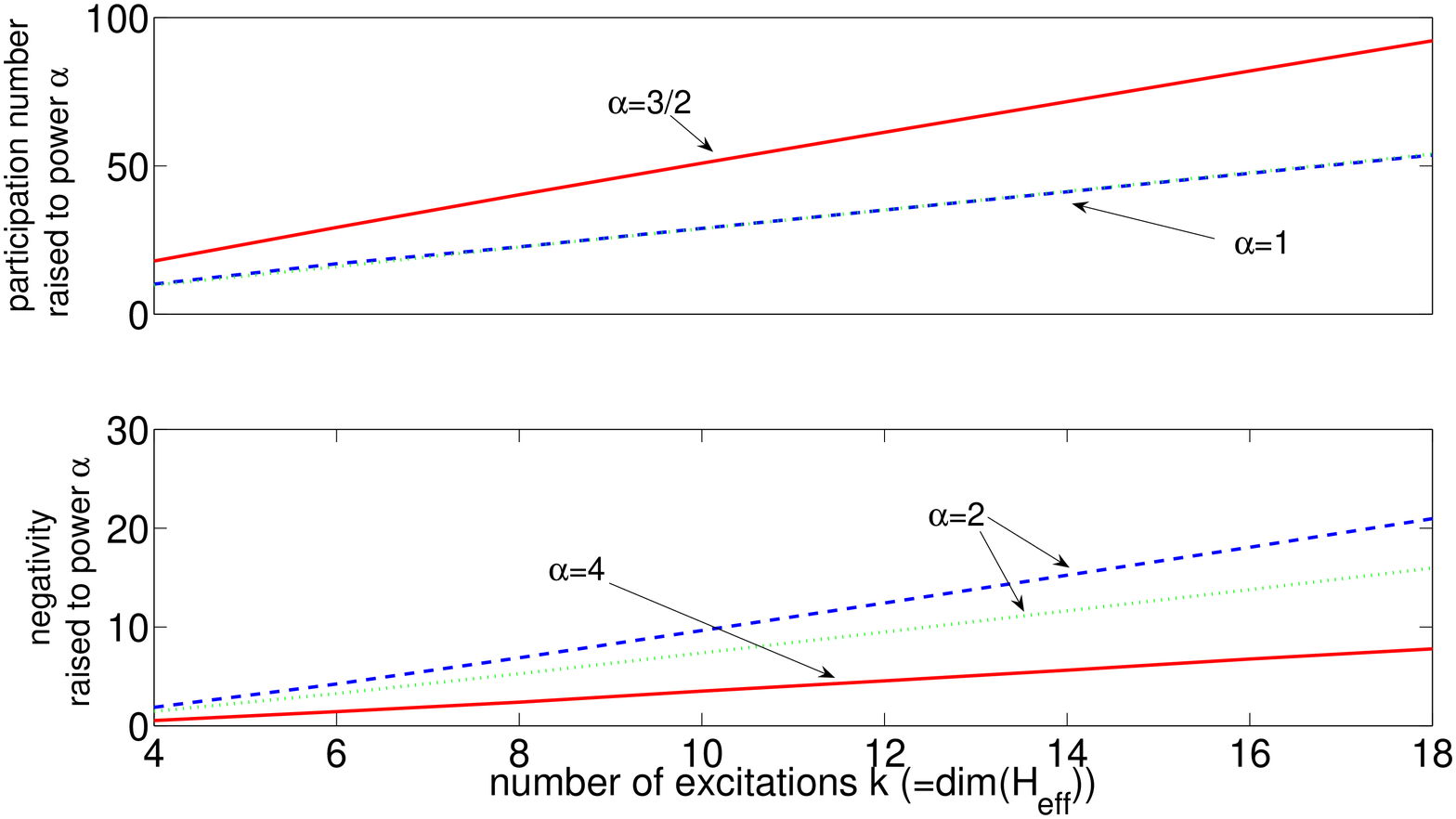, width=14.0cm, clip=} 
\caption{(Color online) Negativity and HS-participation number (raised to various powers to fit linear dependence) measured at the first (in time) maximum on Fig.(\ref{fig:BG1}) for $BG1$ (solid line) and $B1$ (dashed lines) cases and on Fig.(\ref{fig:BGP1}) for $BP1$ (dotted lines) case as a function of the number of excitations $k$ (equal to $\texttt{dim}({\cal H}_{eff})$). Negativity in the $BG1$ case is raised to power $4$  and the negativity in both the $B1$  and $BP1$  cases is raised to power $2$. The  powers of the HS-participation number are $3/2$ in $BG1$ case and $1$ in $B1$ and $BP1$ cases. }
\label{fig:ranil}
\end{figure}

\begin{figure}[t]
\epsfig{file=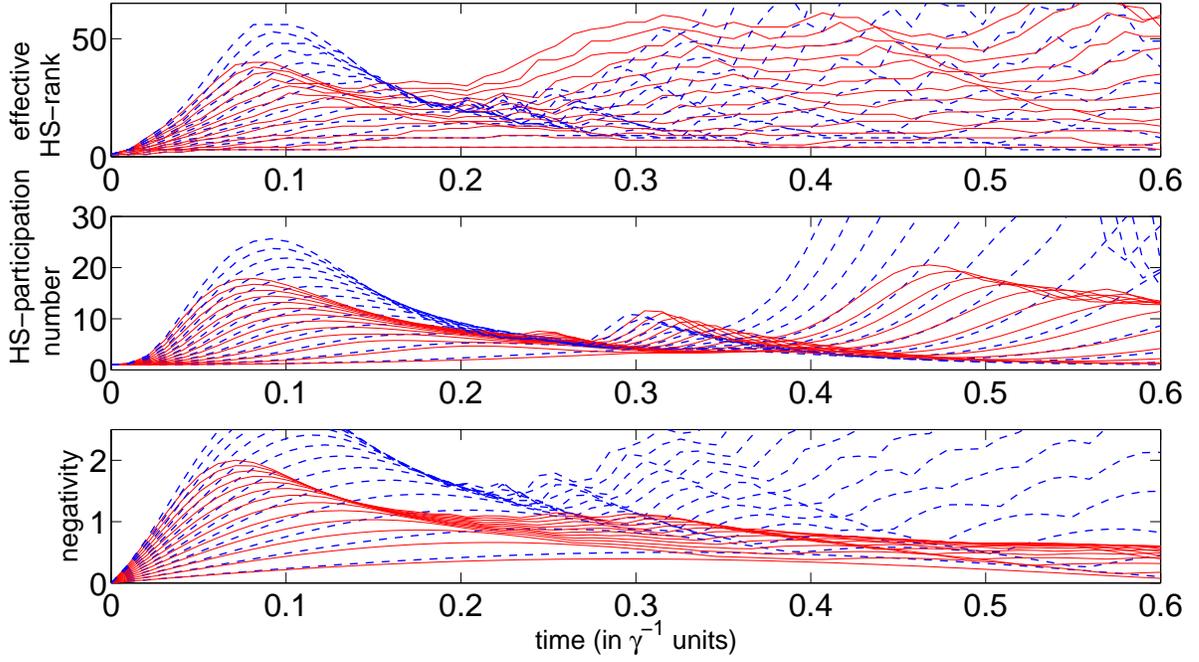, width=18.0cm, clip=} 
\caption{(Color online)The negativity, the effective HS-rank  and the HS-participation number of the  evolving density 
operator: case $BG2$ (solid lines) and $B2$ (dashed lines). In both cases: $2\omega_1=\omega_2=\omega$, in case $BG2$: $\Gamma_1 \omega_1^2=\Gamma_2\omega_2^2=(1/3)\gamma=(1/30)\omega$ and in case $B2$:  $\Gamma_1=\Gamma_2=0$. Initial state $\left|\psi\right\rangle=\left|k \ 0 \right\rangle$ for $k=4,6,...,28$. }
\label{fig:BG2}
\end{figure}

\begin{figure}[t]
\epsfig{file=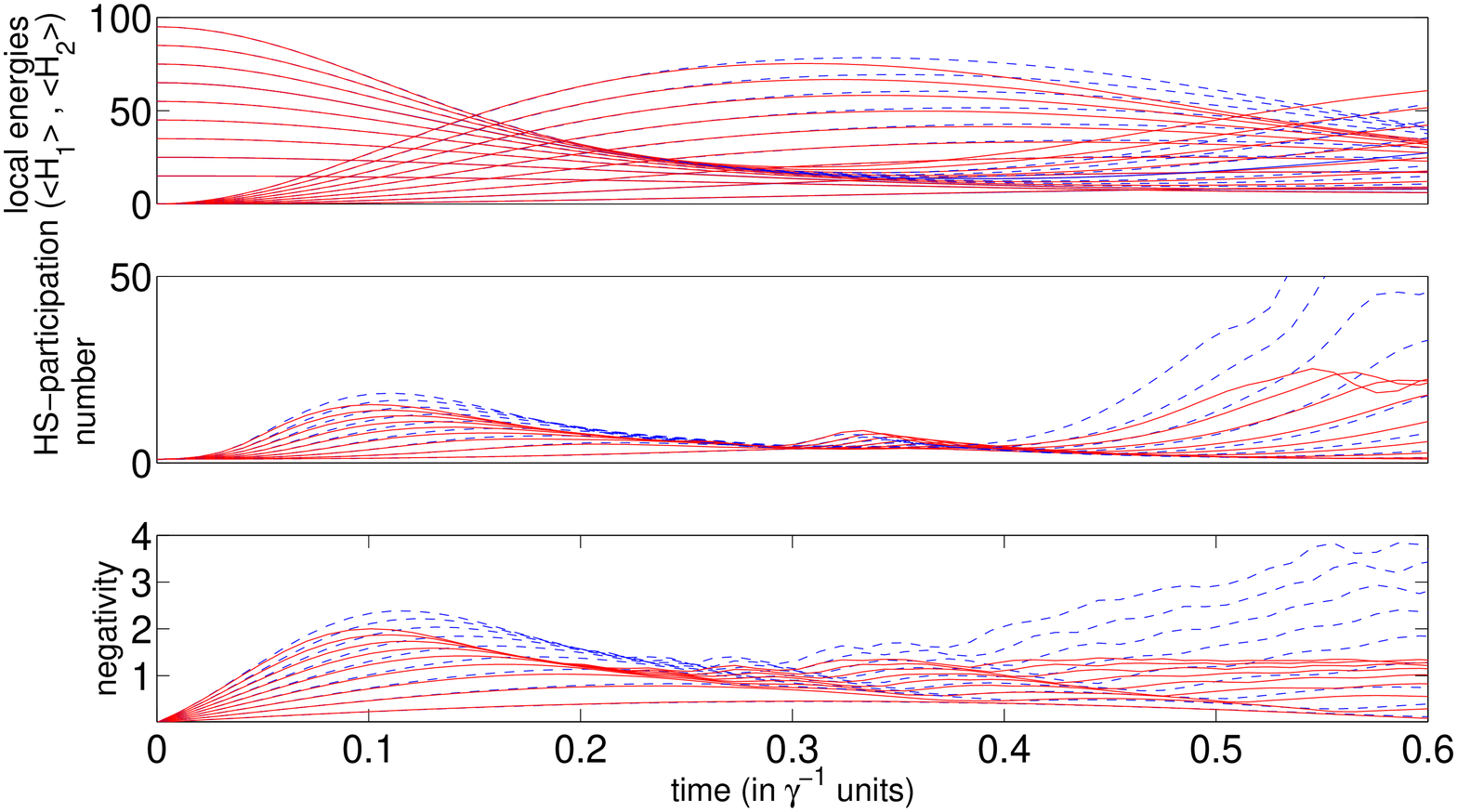, width=18.0cm, clip=} 
\caption{(Color online) The negativity, the local energies  and the HS-participation number of the  density 
operator: cases $BG2$ (solid lines) and $BP2$ (dashed lines). Parameters: $2\omega_1=\omega_2=\omega$ in both cases, $\Gamma_1 \omega_1^2=\Gamma_2\omega_2^2=(1/8)\gamma=(1/80)\omega$ in $BG2$  and $\Gamma_1=\Gamma_2=(1/4)\gamma=(1/40)\omega$, $\phi=2 \pi/7$ in $BP2$ . Initial state $\left|\psi\right\rangle=\left| k \  0\right\rangle$, with $k=4,6,...,20$. }
\label{fig:BGP2}
\end{figure}

\begin{figure}[t]
\epsfig{file=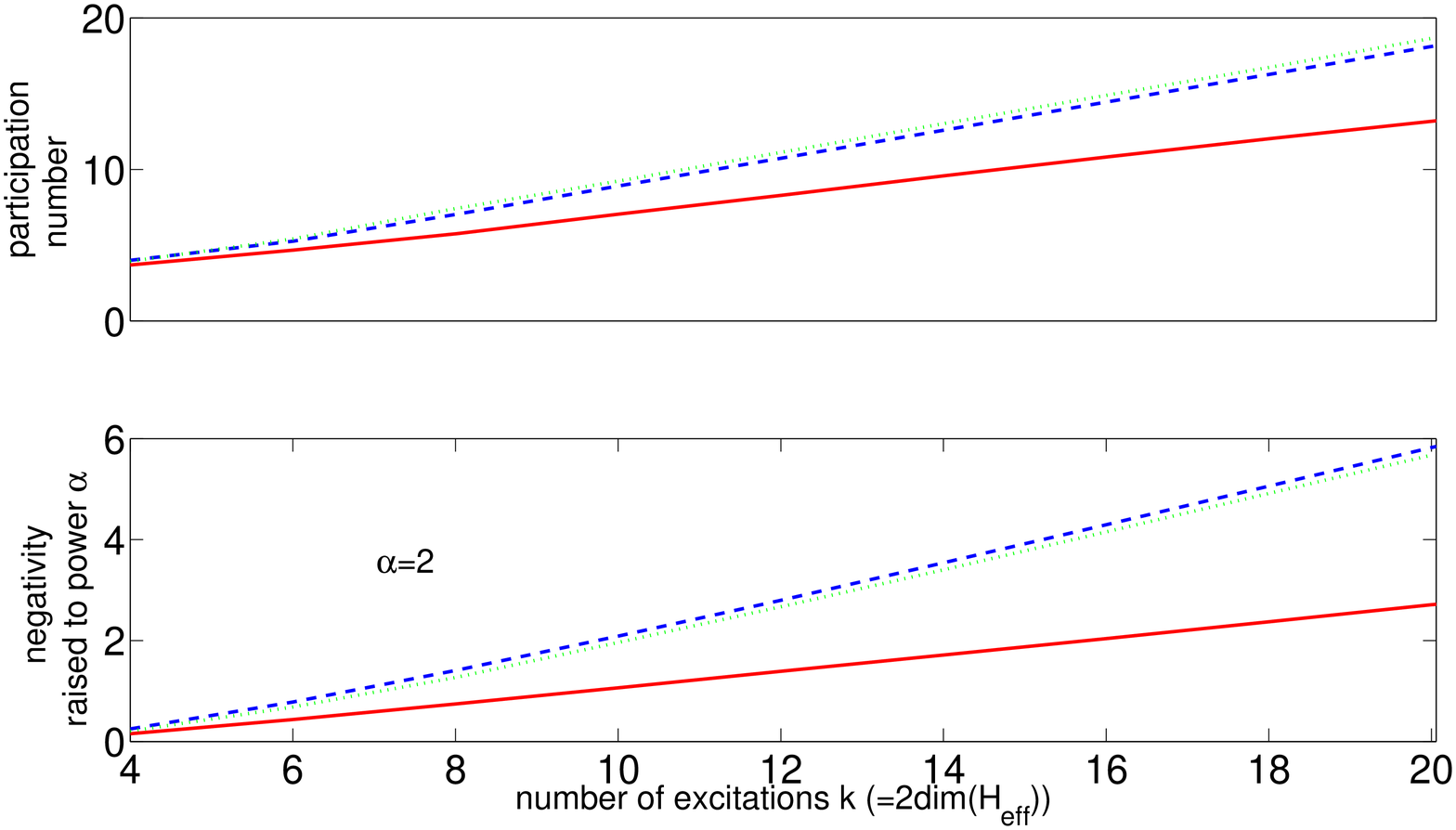, width=14.0cm, clip=} 
\caption{(Color online) Negativity (squared to fit linear dependence) and HS-participation number  measured at the first (in time) maximum on Fig.(\ref{fig:BG2}) for $BG2$ (solid line) and $B2$ (dashed lines) cases and on Fig.(\ref{fig:BGP2}) for $BP2$ (dotted lines) case as a function of the number of excitations $k$ (equal to $\texttt{dim}({\cal H}_{eff})$).}
\label{fig:rani}
\end{figure}

\section{Numerical results}
In the present section the results of numerical calculations of negativity  ${\cal N}(\Op \rho)$, HS-participation number ${\cal \tilde{\kappa}}(\Op \rho)$ and the effective HS-rank  ${\cal \tilde{\chi}}_{0.01}(\Op \rho)$ (Cf. Appendix A) are displayed and
analyzed.  The model is a bipartite composite state of two oscillators, evolving according 
to Eq.(\ref{general}). 
The dynamics simulated is classified according to the type of a local bath, Eq.(\ref{eq:bathtype}), 
and the type of interaction, Eq.(\ref{eq:interaction}):
\begin{itemize}
\item{{\bf AG}) The band-limited interaction ${\cal I}_A$. \textit{Gaussian} pure dephasing (Figs.(\ref{fig:AG},\ref{fig:AGP})).} 
\item{{\bf AP}) The band-limited interaction ${\cal I}_A$ . \textit{Poissonian} pure dephasing (Fig.(\ref{fig:AGP})).}
\item{{\bf A}) The band-limited interaction ${\cal I}_A$ . Isolated reference case (Fig.(\ref{fig:AG})).}
\item{{\bf BG1}) The linear ($r=s=1$) interaction ${\cal I}_B$. \textit{Gaussian} pure dephasing (Figs.(\ref{fig:BG1},\ref{fig:BGP1},\ref{fig:ranil})).}
\item{{\bf BP1}) The linear ($r=s=1$) interaction ${\cal I}_B$. \textit{Poissonian} pure dephasing (Fig.(\ref{fig:BGP1})).}
\item{{\bf B1}) The linear ($r=s=1$) interaction ${\cal I}_B$. Isolated reference case (Figs.(\ref{fig:BG1},\ref{fig:ranil})).}
\item{{\bf BG2}) The nonlinear ($r=1$, $s=2$) interaction ${\cal I}_B$. \textit{Gaussian} pure dephasing (Figs.(\ref{fig:BG2},\ref{fig:rani},\ref{fig:BGP2})).}
\item{{\bf BP2}) The nonlinear ($r=1$,$s=2$) interaction ${\cal I}_B$. \textit{Poissonian} pure dephasing (Fig.(\ref{fig:BGP2})).}
\item{{\bf B2}) The nonlinear ($r=1$,$s=2$) interaction ${\cal I}_B$. Isolated reference case (Figs.(\ref{fig:BG2},\ref{fig:rani})).}
\end{itemize}

In each case the evolution of the composite system starts from a pure uncorrelated state 
$\left|\psi\right\rangle=\left|k 0\right\rangle$, where $k$ is the initial number 
of excitations of the first oscillator, which determines the effective Hilbert 
space dimension of the system. 

\textbf{Case $AG$} (Figs.(\ref{fig:AG},\ref{fig:AGP})). On Fig.(\ref{fig:AG}) the negativity, HS-participation 
number and effective HS-rank of the evolving state  in the presence of the bath
is compared  to the corresponding unitary evolution (case $A$). 
The amplitude of the  negativity in the isolated case grows without bounds 
as the effective Hilbert space dimension $k$ increases. Once the bath is introduced  
the amplitude of the  negativity saturates to a value independent of $k$. 
On the other hand, both the HS-participation number and the effective HS-rank of the evolving state show that the 
total correlations grow without bounds when the effective Hilbert space 
dimension of the system increases. It is interesting to note the qualitative 
difference, most obvious in the unitary evolution (dashed lines), between the dynamics of the HS-participation number and the effective HS-rank on the shorter 
time scale, corresponding to the inverse frequency of the oscillators $\omega^{-1}$. 
While the HS-participation number is smooth on that scale, the effective 
HS-rank displays oscillations which follow closely after 
the corresponding dynamics of the negativity.

\textbf{Case $AP$} Fig.\ref{fig:AGP} compares the dynamics of correlations in case $AP$ to case $AG$. 
The relative strength of couplings  to different types of environments 
is chosen to match the time scales the local energies dephasing in both cases. 
It can be seen that in contrast to case $AG$ both the negativity and the 
HS-participation number  in case $AP$ increase without bounds as the 
effective Hilbert space dimension grows similarly to the corresponding 
unitary evolution displayed on  Fig.(\ref{fig:AG}).

\textbf{Case $BG1$} (Fig.(\ref{fig:BG1},\ref{fig:BGP1},\ref{fig:ranil})). 
On Fig.(\ref{fig:BG1}) the negativity, HS-participation number and effective HS-rank of the evolving state  
in the presence of the bath is compared to the corresponding unitary  
evolution (case $B1$). The amplitude of the  negativity in the isolated case grows 
without bounds as the effective Hilbert space dimension $k$ increases.  
In the bath-on case the amplitude of the  negativity is obviously restricted 
but the quantita
Both HS-participation number and effective HS-rank display the growth of the total correlations 
 without bounds with the effective Hilbert space dimension of the system. 
Note the qualitative difference in dynamics of the two measures.

Fig.(\ref{fig:ranil}) displays the maximal values of the negativity and the HS-participation number obtained in  cases $BG1$, $B1$ and $BP1$ as functions of the effective Hilbert space dimension $k$. It is seen that in the $B1$ case the squared negativity and the HS-participation number scale linearly with $k$ in compliance with the  calculation in Appendix B. The same scaling is found in the case $BP1$. On the other hand, the negativity in the $BG1$ case scales as a fourth root of the effective Hilbert space dimension.
The corresponding HS-participation number measuring the total correlations scales as $k^{2/3}$.

\textbf{Case $BP1$} Fig.(\ref{fig:BGP1}) compares the dynamics of correlations in case $BP1$ to case $BG1$. 
The relative strength of couplings  to different types of environments 
is chosen to match the 
local energies dephasing rates. 
From this figure and Fig.(\ref{fig:ranil}) it can be seen that in contrast to case 
$BG1$ both the negativity and the HS-participation number  
in case $BP1$ follow a dynamical pattern identical to  the corresponding unitary evolution.

\textbf{Case $BG2$} (Fig.(\ref{fig:BG2},\ref{fig:BGP2},\ref{fig:rani})). 
In Fig.(\ref{fig:BG2}) the negativity, HS-participation number and 
effective HS-rank of the evolving state  in the presence of the bath 
is compared to the corresponding unitary  evolution (case $B2$). 
For a nonlinear interaction both the amplitudes and 
time scales of the dynamics depend on the initial state. 
As a consequence, the pattern of behavior changes with the effective Hilbert space dimension. 
This makes it difficult to compare the evolutions corresponding to different $k$. 
Comparing the open to the closed unitary evolutions for a fixed $k$ it is seen that
the global dynamics of the negativity is much stronger effected by the bath
than the dynamics of the total correlations. For example, 
the total correlations may grow in the open dynamics 
similarly to the unitary case, while the negativity at this 
time is decaying in sharp contrast to the corresponding unitary behavior. 

A possible way to compare values of the negativity and the  total correlations at different 
$k$ is to measure the values observed at the first maximum in the evolutions of these quantities. 
These measurement are displayed on  Fig.(\ref{fig:rani}). 
To understand the scaling,  the negativity (squared to fit the linear dependence) and the HS-participation number obtained in cases $BG2$, 
$B2$ and $BP2$ are plotted as a functions of the effective Hilbert space dimension $k$. 
It is found that the negativity   scales  with $\sqrt{k}$ while the 
HS-participation number scale linearly with $k$.

\textbf{Case $BP2$} Fig.(\ref{fig:BGP2}) compares the dynamics of correlations in case $BP2$ to case $BG2$. 
The relative strength of couplings  to different types of environments is chosen to match the 
local energies dephasing rates. The negativity and the HS-participation number  
in case $BP2$  follow a dynamical pattern identical to  the corresponding unitary evolution displayed on  Fig.(\ref{fig:BG2}). 
See also Fig.(\ref{fig:rani}).

\section{Summary and conclusions.}
 Variety of  open interacting bipartite systems were investigated in order to characterize restrictions, imposed by coupling to local environments, on the generation of classical and quantum correlations. 
 
The extent of the generated quantum correlations is determined by the interplay of two competing forces: the interaction, leading to development of entanglement, and the local decoherence, inducing a decay of entanglement.
The relative magnitudes of the local decoherence rates and the cut-off frequency of the interaction in the effective Hilbert space of the composite system determines the relative size of decoherence- and interaction-dominated regions of the density operator in local robust states basis. The presence of the decoherence-dominated regions constrains the structure of the evolving composite density operator, restricting the extent of entanglement, generated by the interaction.

The character of restriction depends on the type of  bath and the type of the interaction. The two different praradigms of the dephasing, the Poissonian and the Gaussian, lead to very different correlation dynamics. In models with band-limited decoherence such as the Poissonian pure dephasing model, either the  decoherence  or the  interaction dominates the dynamics, depending on the relataive strength of the coupling constants and irrespectively of initial state.   Numerical calculations performed on a bipartite system of two interacting harmonic oscillators, coupled to local  Poissonian baths, support this conclusion.

Open systems with Gaussian pure dephasing belong to a different class of  models. This class is characterized by unbounded growth of the decoherence time scales with the effective Hilbert space dimension of the system. As a consequence,  constrains on the structure of evolving state and restriction on the extent of entanglement are generally expected.  Still the precise character of the restriction depends on the type of interaction between the subsystems. 
Coupling  local Gaussian  environments to subsystems with band-limited interaction between them, imposes an upper bound on the extent of generated entanglement, which is independent of the effective Hilbert space dimension of the system. As a consequence, asymptotically, i.e. at sufficiently large effective dimension, the generated entanglement is negligible, compared to entanglement generated in the corresponding unitary dynamics. 
Interactions which are not band-limited  generally  produce extensive entanglement, notwithstanding the type of local environment. Nonetheless, in models with local Gaussian  environments the scaling  of entanglement with the effective dimension is limited by the  local decoherence. The precise limit depends on the nonlinearity of the interaction. In the model of two nonlinearly interacting harmonic oscillators stronger nonlinearity implies weaker bounds on the generated entanglement. When the nonlinearity exceeds some maximal value no restriction on the extent of  entanglement is expected. Numerical calculations  support these predictions.

Estimation of bounds on negativity in the evolving state was based on analysis of the structure of the density matrix in particular local robust states bases.  Relating the negativity to the structure of the density operator was facilitated  by the observation that the evolving states are Schmidt-correlated  due to particular conservation laws observed by the interactions. The corresponding Schmidt bases are built of local robust states selected by local purely dephasing environments $-$ the local energy bases. Since the presence of exact conservation laws is nongeneric in  physical models, it should be noted that numerical evidence shows that the qualitative picture presented above is robust.

Dynamics of the total correlations was investigated numerically to compare with the corresponding dynamics of the entanglement. It was found that evolution of the total (and, as a consequence, classical) correlations display a different dynamical pattern.  In the band-limited interaction model, the amplitude of the total correlations grows without bounds with the effective Hilbert space dimension, while the negativity tends to an asymptotic behavior independent of the effective dimension. In the linear interaction model, though the amplitudes of both the quantum and the total correlations  grow without bounds with the effective Hilbert space dimension, the total correlations scale with a higher power of the dimension. In the nonlinear interaction, comparison is impeded by the fact that the evolution of both the entanglement and the total correlations display a variety of time-scales. Nonetheless, inspection of the numerical evidence shows that the  total correlations always scale with a higher power of the effective Hilbert space dimension. These findings can be informally interpreted as a trade-off between the classical and quantum correlations: since the total correlations are (relatively) unaffected by the environment, restriction on the entanglement generation must be "compensated" by the growth of the classical correlations.

Considering the restriction on the generation of entanglement, a natural question arises: is the observed restriction substantial, i.e. is the given partition of the composite system  meaningful? When can a composite systems be regarded as \textit{approximately disentangled}?  The answer depends on the definition of the relevant scale of a measure  of entanglement  in the evolving system. Is the scale unity or some power of the effective Hilbert space dimension or neither? 

One possibility is to compare the entanglement, generated in the open evolution to the entanglement, generated in the corresponding unitary evolution. Numerical evidence obtained in the present study shows that entanglement is always relatively restricted in the open system dynamics. In some cases, such as the Gaussian pure dephasing, it can even be negligible in asymptotically large Hilbert space dimensions. This comparison elucidates the role of the decoherence in constraining the generation of the quantum correlations. Nevertheless, the magnitude of the entanglement generated in a particular open evolution may still be large in some absolute sense. 

An alternative scale of entanglement is set by  the maximal entanglement compatible with the effective Hilbert space dimension. The results of the present study show that in some  models, such as  the Gaussian pure dephasing and weakly nonlinear or band-limited interactions, coupling to local environments does the job, i.e. it restricts the generated entanglement to bounds, negligible compared to the maximal compatible entanglement. Still, in all cases apart from a band-limited type of the interaction, entanglement, generated on the interaction time scale in the open system evolution,  grows without bounds with the effective Hilbert space dimension. As a consequence, this scale may become irrelevant in large effective Hilbert dimensions, due to a limited experimental resolution.
 
To conclude, common models of local decoherence do not  provide a universal pathway to an approximately disentangled evolution of a  bipartite composite system in the presence of interaction. 
It follows that, contrary to expectations,  coupling to local environments does not generally validate partition of composite quantum systems.

\appendix

\section{The Schmidt rank and the HS-Schmidt rank.}

The definition of the Schmidt rank of the bipartite composite state \cite{schmidt,peres} is reviewed.
Let $\left|\psi\right\rangle$ be a state in the composite Hilbert space 
${\cal H}_{12}={\cal H}_1 \otimes {\cal H}_2$. 
There exist a following representation (a Schmidt decomposition ) of the state: 
$\left|\psi\right\rangle=\sum_i c_i \left|i\right\rangle_1\otimes\left|i\right\rangle_2$, 
where $\left|i\right\rangle_{1,2}$  is an orthonormal basis in the  Hilbert 
space ${\cal H}_{1,2}$. While a Schmidt decomposition is not unique, 
the set of non vanishing coefficients $c_i$ is invariant (modulo  
irrelevant phases) under the local unitary transformations 
and is characteristic of the state $\left|\psi\right\rangle$. 
This set is shown to be the square root of the spectrum of the reduced 
density operator of either subsystem. The number of non vanishing coefficients $c_i$ is called the 
Schmidt rank $\chi(\psi)$ of the state and
equals the rank of the reduced density operator of either subsystem: $\chi(\psi)=\texttt{rank}\{\texttt{Tr}_1\{\Op \rho_{12}\} \}$.  To calculate the Schmidt rank 
of a state expressed in an arbitrary tensor product basis 
$\left|\psi \right\rangle=\sum_{ij}c_{ij}\left|i \right\rangle_1\left|j\right\rangle_2$ one calculates the rank of the matrix $\rho_2=C^{\dagger}C$, where $C_{ij}=c_{ij}$: 
$\texttt{Tr}_1\{\Op \rho_{12}\}=\sum_{\{n,i,j,k,l\}}c_{ij}c^*_{kl}\delta_{in}\delta_{kn}\left|j\right\rangle\left\langle l\right|=\sum_{\{i,j,l\}}c_{ij}c^*_{il}\left|j\right\rangle\left\langle l\right|=\sum_{\{j,l\}}(\rho_2)_{lj}\left|j\right\rangle\left\langle l\right|$.

The Schmidt rank characterizes the extent of correlations present in the state.
The uncorrelated (product) state has $\chi=1$ but generally $\chi(\psi) \le \min\left\{\dim({\cal H}_1), \dim({\cal H}_2) \right\}$. 
The maximally correlated state has $\chi(\psi) = \min\left\{\dim({\cal H}_1), \dim({\cal H}_2) \right\}$ and $c_i=c_j$, 
$\forall i,j$. 
Generally, some of the coefficients $c_i$ are much smaller than others and as a consequence dropping the corresponding 
contributions to the Schmidt decomposition does not lead to an observable effect. 
This suggests a definition of the physically reasonable \textit{effective Schmidt rank} \cite{vidal2} $\chi_{\epsilon}$: 
$\chi_{\epsilon}(\psi)\equiv \chi(\psi')$, with $\left|\psi'\right\rangle=\sum_{i\in I_{\epsilon}} c_i\left|i\right\rangle_1\otimes\left|i\right\rangle_2$, 
where $I_{\epsilon}$ is the smallest set of indices such that 
$\left\| \left|\psi\right\rangle-\left|\psi'\right\rangle \right\|< \epsilon$. 
An alternative measure is a participation number \cite {Eberly} 
$\kappa (\psi)\equiv 1/ Tr\{\Op \rho_2^2 \}$ with $\Op \rho_2=\texttt{Tr}_1\{\Op \rho_{12}\}$. 
The participation number of a state, characterized by $M$ equal substantial contributions 
to its Schmidt decomposition is seen to be  $M$, which motivates the definition.  

A mixed state displays both quantum (entanglement) and classical correlations. The extent of the \textit{total} correlations can be characterized by the Schmidt rank of a {density operator}. With a slight abuse of terminology the term  HS-Schmidt rank (HS indicating the Hilbert-Schmidt space) or just HS-rank is adopted. The definition of the HS-rank views the density operator of 
a composite system as a (unnormalized) pure state ("superket" \cite{zwolak}) in the  Hilbert-Schmidt space of system operators. The  Schmidt rank of the corresponding "superket" defines the HS-rank (denoted $\tilde{\chi}(\Op \rho)$) of the density operator. The notions of the effective Schmidt rank $\chi_{\epsilon}(\psi)$ and the participation number $\kappa(\psi)$ can be  transfered to the HS-rank of the density operator. For brevity, the corresponding measures of the total correlations are termed effective HS-rank and HS-participation number and denoted by $\tilde{\chi}_{\epsilon}(\Op \rho)$ and  $\tilde{\kappa}(\Op \rho)$, respectively.

The calculation of the HS-rank proceeds as follows.
Let $\Op \rho_{12}=\sum_{\{i,j,k,l\}}\rho_{ijkl}\left|ij\right\rangle\left\langle kl\right|$ 
be a density operator of the composite system. In the superket notation it has the form 
$\left|\Op \rho\right\rangle_{12}=\sum_{\{i,j,k,l\}}\rho_{ijkl}\left|\left|ij\right\rangle\left\langle kl\right|\right\rangle$. 
The corresponding density superoperator is 
${\cal R}_{12}(\Op \rho)=\sum_{\{i,j,k,l,i',j',k',l'\}}\rho_{ijkl}\rho^*_{i'j'k'l'}\left|\left|ij\right\rangle\left\langle kl\right|\right\rangle   \left\langle \left|i'j'\right\rangle\left\langle k'l'\right|\right|$ 
and the reduced density superoperator is ${\cal R}_{2}(\Op \rho)=\texttt{Tr}_1\{{\cal R}_{12}(\Op \rho)\}=\sum_{\{i,j,k,l,i',j',k',l',m,n\}}\rho_{ijkl}\rho^*_{i'j'k'l'}\delta_{mi}\delta_{nk}\delta_{mi'}\delta_{nk'}\left|\left|j\right\rangle\left\langle l\right|\right\rangle   \left\langle \left|j'\right\rangle\left\langle l'\right|\right|=\sum_{\{j,l,j',l'\}}R_{jlj'l'}\left|\left|j\right\rangle\left\langle l\right|\right\rangle     \left\langle \left|j'\right\rangle\left\langle l'\right|\right|$, where  $R_{jlj'l'}=\sum_{ik}\rho_{ijkl}\rho^*_{i'j'k'l'}$. 
The HS-rank $\tilde{\chi}(\Op \rho)$ of the density operator $\Op \rho$ is
$\tilde{\chi}(\Op \rho)=\texttt{rank}\{{\cal R}_{2}(\Op \rho) \}$. The effective HS-rank and the HS-participation number are calculated similarly.

Finally, note that $\tilde{\chi}(\left|\psi\right\rangle\left\langle \psi\right|)=\chi(\psi)^2$. 
In fact,  $\chi_{\psi}^2=(\texttt{rank}\{{\Op \rho}_{2} \})^2=\texttt{rank}\{{\Op \rho}_{2}\otimes {\Op \rho}_{2}^T \}=\texttt{rank}\{\sum_{\{i,j,l,k,j',l'\}}a_{ij}a^*_{il}a_{kj'}a^*_{kl'}\left|j\right\rangle\left\langle l\right|\otimes\left|l'\right\rangle\left\langle j'\right|\}=\texttt{rank}\{\sum_{\{i,j,l,k,j',l'\}}\rho_{ijkl'}\rho^*_{ilkj'}\left|j\right\rangle\left\langle l\right|\otimes\left|l'\right\rangle\left\langle j'\right|\}=\texttt{rank}\{\sum_{\{j,l,j',l'\}}R_{jl'lj'}\left|j\right\rangle\left\langle l\right|\otimes\left|l'\right\rangle\left\langle j'\right|\}=\texttt{rank}\{\sum_{\{j,l,j',l'\}}R_{jlj'l'}\left|j\right\rangle\left\langle j'\right|\otimes\left|l\right\rangle\left\langle l'\right|\}=\texttt{rank}\{\sum_{\{j,l,j',l'\}}R_{jlj'l'}\left|j l\right\rangle\left\langle j'l'\right|\}=\tilde{\chi}(\left|\psi\right\rangle\left\langle \psi\right|)$.

\section{Calculation of the   effective Schmidt rank  of the composite state of two linearly interacting harmonic oscillators}

A system of two linearly interacting harmonic oscillators is considered with the Hamiltonian 
$\Op H=\omega (\Op a_{1}^{\dagger}\Op a_{1}+\Op a_{2}^{\dagger}\Op a_{2})+\gamma (\Op a_{1}^{\dagger}\Op a_{2}+\Op a_{2}^{\dagger}\Op a_{1})$  
The initial state is  $\left|\psi(0)\right\rangle=\left|0 k\right\rangle$ in the local energies basis.
The state at $t>0$ becomes: 
$\left|\psi(t)\right\rangle=\sum_{n=0}^k c_n \left|  k-n \ n \right\rangle$, where 
$c_n(t)=\sqrt{\frac{k!\cos(\gamma t)^{2n}\sin(\gamma t)^{2(k-n)}}{n!(k-n)!}}e^{-i\omega k t}$. 
The  width $\Delta_k $ of the distribution of expansion coefficients $|c_n|^2$ is estimated at $t=\pi/4\gamma$ for  $k\gg 1$. 
This width is  a reasonable estimate for the amplitude of the effective Schmidt rank of the state: $\chi (\psi)\approx \Delta_k$. 
 
The distribution of the coefficients  $|c_n(\pi/4\gamma)|^2=\frac{k!}{2^k n!(k-n)!}$ 
is peaked around $n=k/2$. To estimate $\Delta $ it is assumed that $\Delta_k  \ll k$. 
$\Delta_k $ is defined by: $\frac{\partial^2}{\partial n^2} |c_n(\pi/4\gamma)|^2|_{n=n^*}=0$, 
where $n^*=k/2-\Delta_k /2$. Performing the derivation under the Stirling approximation for 
the factorials (valid at $k \gg 1$) leads to $\frac{k}{n^*(k-n^*)}=\ln^2(\frac{k-n^*}{n^*})$. 
For highly peaked distribution  $\frac{k-n^*}{n^*}-1\ll 1$, 
therefore $\ln^2(\frac{k-n^*}{n^*})\approx (\frac{k-2n^*}{n^*})^2$. 
Also $\frac{k}{n^*(k-n^*)}\approx \frac{4}{k}$ to the leading order in $\frac{k-2n^*}{n^*}$. 
Finally  $\frac{4}{k} \approx (\frac{k-2n^*}{n^*})^2 \approx (\frac{k-2n^*}{k/2})^2=(\frac{\Delta_k }{k/2})^2$ 
from which $\Delta_k=\sqrt{k}$ and $\chi (\psi)\approx \Delta_k=\sqrt{k}$. As follows from the relation $\tilde{\chi}(\left|\psi\right\rangle\left\langle \psi\right|)=\chi(\psi)^2$, proved in Appendix A, the amplitude of the effective HS-rank scales as $k$.

The obtained result can be used to estimate the amplitude of the negativity in the pure state evolution. 
In fact ${\cal N}(\left|\psi(t)\right\rangle\left\langle \psi(t)\right|)=\frac{1}{2}\left( |\sum_n c_n|^2-1   \right)$ by Ref.\cite{vidal}. Taking $c_n=\frac{1}{\sqrt{\Delta_k }}=\frac{1}{\sqrt[4]{k}}$ for the purpose of scaling we obtain ${\cal N}(\left|\psi(\pi/4\gamma)\right\rangle\left\langle \psi(\pi/4\gamma)\right|)=\frac{1}{2}\sqrt{k}$ for the amplitude of the negativity.

\begin{acknowledgments}We are grateful to L. Diosi, D. Steinitz and Y. Shimoni for useful comments.
Work supported by Binational US-Israel Science Foundation (BSF).
The Fritz Haber Center is supported
by the Minerva Gesellschaft f\"{u}r die Forschung GmbH M\"{u}nchen, Germany.
\end{acknowledgments}

%\bibliography{/usr/people/ronnie15/ronnie/Text/Michael/Anza/Anz4/pumpprob,/usr/people/ronnie15/ronnie/Text/Database/pub} 
   %\bibliography{Z:/texs/pumpprob,Z:/texs/Database/pub} 
%\bibliography{C:/Tex/pumpprob,C:/Tex/Database/pub} 

\begin{thebibliography}{63}
\expandafter\ifx\csname natexlab\endcsname\relax\def\natexlab#1{#1}\fi
\expandafter\ifx\csname bibnamefont\endcsname\relax
  \def\bibnamefont#1{#1}\fi
\expandafter\ifx\csname bibfnamefont\endcsname\relax
  \def\bibfnamefont#1{#1}\fi
\expandafter\ifx\csname citenamefont\endcsname\relax
  \def\citenamefont#1{#1}\fi
\expandafter\ifx\csname url\endcsname\relax
  \def\url#1{\texttt{#1}}\fi
\expandafter\ifx\csname urlprefix\endcsname\relax\def\urlprefix{URL }\fi
\providecommand{\bibinfo}[2]{#2}
\providecommand{\eprint}[2][]{\url{#2}}

\bibitem[{\citenamefont{{N. Moiseyev}}(1983)}]{Moiseyev83}
\bibinfo{author}{\bibnamefont{{N. Moiseyev}}}, \bibinfo{journal}{Chem. Phys.
  Lett.} \textbf{\bibinfo{volume}{98}}, \bibinfo{pages}{233}
  (\bibinfo{year}{1983}).

\bibitem[{\citenamefont{{P. Jungwirth, M. Roeselova, R.B.
  Gerber}}(1999)}]{Jungwirth99}
\bibinfo{author}{\bibnamefont{{P. Jungwirth, M. Roeselova, R.B. Gerber}}},
  \bibinfo{journal}{J. Chem. Phys.} \textbf{\bibinfo{volume}{110}},
  \bibinfo{pages}{9833} (\bibinfo{year}{1999}).

\bibitem[{\citenamefont{Peres}(1998)}]{peres}
\bibinfo{author}{\bibfnamefont{A.}~\bibnamefont{Peres}},
  \emph{\bibinfo{title}{{Quantum Theory: Concepts and Methods}}}
  (\bibinfo{publisher}{Kluwer, Boston}, \bibinfo{year}{1998}).

\bibitem[{\citenamefont{{R. F. Werner}}(1989)}]{werner}
\bibinfo{author}{\bibnamefont{{R. F. Werner}}}, \bibinfo{journal}{Phys. Rev. A}
  \textbf{\bibinfo{volume}{40}}, \bibinfo{pages}{4277} (\bibinfo{year}{1989}).

\bibitem[{\citenamefont{{W.H. Miller}}(2002)}]{Miller02}
\bibinfo{author}{\bibnamefont{{W.H. Miller}}}, \bibinfo{journal}{J. Phys.
  Chem.} \textbf{\bibinfo{volume}{106}}, \bibinfo{pages}{8132}
  (\bibinfo{year}{2002}).

\bibitem[{\citenamefont{{E.J. Heller}}(2006)}]{Heller06}
\bibinfo{author}{\bibnamefont{{E.J. Heller}}}, \bibinfo{journal}{Acc. Chem.
  Res.} \textbf{\bibinfo{volume}{39}}, \bibinfo{pages}{127}
  (\bibinfo{year}{2006}).

\bibitem[{\citenamefont{{M. H. Beck, A.Jackle, G.A.Worth,
  H.D.Meyer}}(2000)}]{Meyer00}
\bibinfo{author}{\bibnamefont{{M. H. Beck, A.Jackle, G.A.Worth, H.D.Meyer}}},
  \bibinfo{journal}{Phys. Rep.} \textbf{\bibinfo{volume}{324}},
  \bibinfo{pages}{1} (\bibinfo{year}{2000}).

\bibitem[{\citenamefont{{G. Vidal}}(2003)}]{vidal1}
\bibinfo{author}{\bibnamefont{{G. Vidal}}}, \bibinfo{journal}{Phys. Rev. Lett.}
  \textbf{\bibinfo{volume}{91}}, \bibinfo{pages}{147902}
  (\bibinfo{year}{2003}).

\bibitem[{\citenamefont{{G. Vidal}}(2004)}]{vidal2}
\bibinfo{author}{\bibnamefont{{G. Vidal}}}, \bibinfo{journal}{Phys. Rev. Lett.}
  \textbf{\bibinfo{volume}{93}}, \bibinfo{pages}{040502}
  (\bibinfo{year}{2004}).

\bibitem[{\citenamefont{{M. Zwolak and G. Vidal}}(2004)}]{zwolak}
\bibinfo{author}{\bibnamefont{{M. Zwolak and G. Vidal}}},
  \bibinfo{journal}{Phys. Rev. Lett.} \textbf{\bibinfo{volume}{93}},
  \bibinfo{pages}{207205} (\bibinfo{year}{2004}).

\bibitem[{\citenamefont{{Y.Y.Shi, L.M.Duan and
  G.Vidal}}(quant-ph/0511070)}]{vidal06}
\bibinfo{author}{\bibnamefont{{Y.Y.Shi, L.M.Duan and G.Vidal}}}
  (\bibinfo{year}{quant-ph/0511070}).

\bibitem[{\citenamefont{Jozsa and Linden}(2003)}]{Jozsa}
\bibinfo{author}{\bibfnamefont{R.}~\bibnamefont{Jozsa}} \bibnamefont{and}
  \bibinfo{author}{\bibfnamefont{N.}~\bibnamefont{Linden}},
  \bibinfo{journal}{Proc. R. Soc. Lond. A} \textbf{\bibinfo{volume}{459}},
  \bibinfo{pages}{2011} (\bibinfo{year}{2003}).

\bibitem[{\citenamefont{{D. Giulini, E. Joos, C. Kiefer, J. Kupsch, I.O.
  Stamatescu and H.D. Zeh (Eds.)}}(2003)}]{Joos}
\bibinfo{author}{\bibnamefont{{D. Giulini, E. Joos, C. Kiefer, J. Kupsch, I.O.
  Stamatescu and H.D. Zeh (Eds.)}}}, \emph{\bibinfo{title}{Decoherence and the
  Appearance of a Classical World in Quantum Theory}}
  (\bibinfo{publisher}{Springer, Berlin, Heidelberg}, \bibinfo{year}{2003}).

\bibitem[{\citenamefont{{M. A. Nielsen and I. L. Chuang}}(2000)}]{Nielsen}
\bibinfo{author}{\bibnamefont{{M. A. Nielsen and I. L. Chuang}}},
  \emph{\bibinfo{title}{Quantum Computation and Quantum Information}}
  (\bibinfo{publisher}{Cambridge University Press}, \bibinfo{year}{2000}).

\bibitem[{\citenamefont{Braun}(2002)}]{Braun}
\bibinfo{author}{\bibfnamefont{D.}~\bibnamefont{Braun}},
  \bibinfo{journal}{Phys. Rev. Lett.} \textbf{\bibinfo{volume}{89}},
  \bibinfo{pages}{277901} (\bibinfo{year}{2002}).

\bibitem[{\citenamefont{{F. Benatti, R. Floreanini and M.
  Piani}}(2003)}]{Benatti}
\bibinfo{author}{\bibnamefont{{F. Benatti, R. Floreanini and M. Piani}}},
  \bibinfo{journal}{Phys. Rev. Lett.} \textbf{\bibinfo{volume}{91}},
  \bibinfo{pages}{070402} (\bibinfo{year}{2003}).

\bibitem[{\citenamefont{{P.J. Dodd and J.J. Halliwell}}(2004)}]{Halliwell04}
\bibinfo{author}{\bibnamefont{{P.J. Dodd and J.J. Halliwell}}},
  \bibinfo{journal}{Phys. Rev. A} \textbf{\bibinfo{volume}{69}},
  \bibinfo{pages}{052105} (\bibinfo{year}{2004}).

\bibitem[{\citenamefont{{P.J. Dodd }}(2004)}]{Dodd04}
\bibinfo{author}{\bibnamefont{{P.J. Dodd }}}, \bibinfo{journal}{Phys. Rev. A}
  \textbf{\bibinfo{volume}{69}}, \bibinfo{pages}{052106}
  (\bibinfo{year}{2004}).

\bibitem[{\citenamefont{{T. Yu and J. H. Eberly}}(2003)}]{Eberly03}
\bibinfo{author}{\bibnamefont{{T. Yu and J. H. Eberly}}},
  \bibinfo{journal}{Phys. Rev. B} \textbf{\bibinfo{volume}{68}},
  \bibinfo{pages}{165322} (\bibinfo{year}{2003}).

\bibitem[{\citenamefont{{T. Yu and J. H. Eberly}}(2006)}]{Eberly06}
\bibinfo{author}{\bibnamefont{{T. Yu and J. H. Eberly}}},
  \bibinfo{journal}{Opt. Comm.} \textbf{\bibinfo{volume}{264}},
  \bibinfo{pages}{393} (\bibinfo{year}{2006}).

\bibitem[{\citenamefont{{M. Franca Santos, P. Milman, L. Davidovich and N.
  Zagury}}(2006)}]{FrancaSantos}
\bibinfo{author}{\bibnamefont{{M. Franca Santos, P. Milman, L. Davidovich and
  N. Zagury}}}, \bibinfo{journal}{Phys. Rev. A} \textbf{\bibinfo{volume}{73}},
  \bibinfo{pages}{040305(R)} (\bibinfo{year}{2006}).

\bibitem[{\citenamefont{Diosi}(2003)}]{diosi03}
\bibinfo{author}{\bibfnamefont{L.}~\bibnamefont{Diosi}},
  \emph{\bibinfo{title}{Progressive Decoherence and Total Environmental
  Disentanglement}} in \emph{\bibinfo{title}{Irreversible Quantum Dynamics}} (\bibinfo{publisher}{Springer, Berlin, Heidelberg},
  \bibinfo{address}{edited by F. Benatti and R. Floreanini},
  \bibinfo{year}{2003}).
  
\bibitem[{\citenamefont{{T. Yu and J. H. Eberly}}(2002)}]{Eberly02}
\bibinfo{author}{\bibnamefont{{T. Yu and J. H. Eberly}}},
  \bibinfo{journal}{Phys. Rev. B} \textbf{\bibinfo{volume}{66}},
  \bibinfo{pages}{193306} (\bibinfo{year}{2002}).

\bibitem[{\citenamefont{{T. Yu and J. H. Eberly}}(2004)}]{Eberly04}
\bibinfo{author}{\bibnamefont{{T. Yu and J. H. Eberly}}},
  \bibinfo{journal}{Phys. Rev. Lett.} \textbf{\bibinfo{volume}{93}},
  \bibinfo{pages}{140404} (\bibinfo{year}{2004}).

\bibitem[{\citenamefont{{A. R. R. Carvalho,F. Mintert, S. Palzer and A.
  Buchleitner}}(2004)}]{Buchleitner04}
\bibinfo{author}{\bibnamefont{{A. R. R. Carvalho,F. Mintert, S. Palzer and A.
  Buchleitner}}}, \bibinfo{journal}{Phys. Rev. Lett.}
  \textbf{\bibinfo{volume}{93}}, \bibinfo{pages}{230501}
  (\bibinfo{year}{2004}).

\bibitem[{\citenamefont{Dur and Briegel}(2004)}]{Dur}
\bibinfo{author}{\bibfnamefont{W.}~\bibnamefont{Dur}} \bibnamefont{and}
  \bibinfo{author}{\bibfnamefont{H.~J.} \bibnamefont{Briegel}},
  \bibinfo{journal}{Phys. Rev. Lett.} \textbf{\bibinfo{volume}{92}},
  \bibinfo{pages}{180403} (\bibinfo{year}{2004}).

\bibitem[{\citenamefont{{A.R.R. Carvalho,F. Mintert, S. Palzer and A.
  Buchleitner}}(quant-ph/0508114)}]{Buchleitner}
\bibinfo{author}{\bibnamefont{{A.R.R. Carvalho,F. Mintert, S. Palzer and A.
  Buchleitner}}} (\bibinfo{year}{quant-ph/0508114}).

\bibitem[{\citenamefont{{F. Mintert, A. R.R. Carvalho, M. Kus and A.
  Buchleitner}}(2005)}]{Mintert}
\bibinfo{author}{\bibnamefont{{F. Mintert, A. R.R. Carvalho, M. Kus and A.
  Buchleitner}}}, \bibinfo{journal}{Phys. Rep.} \textbf{\bibinfo{volume}{415}},
  \bibinfo{pages}{207} (\bibinfo{year}{2005}).

\bibitem[{\citenamefont{{S.B. Lee, J. B. Xu }}(2003)}]{Lee}
\bibinfo{author}{\bibnamefont{{S.B. Lee, J. B. Xu }}},
  \bibinfo{journal}{Physics Letters A} \textbf{\bibinfo{volume}{311}},
  \bibinfo{pages}{313} (\bibinfo{year}{2003}).

\bibitem[{\citenamefont{M.~B.~Plenio and Eisert}(2004)}]{Plenio04}
\bibinfo{author}{\bibfnamefont{J.}~\bibnamefont{M.~B.~Plenio}}
  \bibnamefont{and} \bibinfo{author}{\bibfnamefont{J.}~\bibnamefont{Eisert}},
  \bibinfo{journal}{New J. Phys.} \textbf{\bibinfo{volume}{6}},
  \bibinfo{pages}{36} (\bibinfo{year}{2004}).

\bibitem[{\citenamefont{Abragam}(1961)}]{Abragam}
\bibinfo{author}{\bibfnamefont{A.}~\bibnamefont{Abragam}},
  \emph{\bibinfo{title}{{The Principles of Nuclear Magnetism}}}
  (\bibinfo{publisher}{Clarendon Press}, \bibinfo{address}{Oxford},
  \bibinfo{year}{1961}).

\bibitem[{\citenamefont{C.W.Gardiner}(1983)}]{Gardiner}
\bibinfo{author}{\bibnamefont{C.W.Gardiner}}, \emph{\bibinfo{title}{{Handbook
  of Stochastic Methods}}} (\bibinfo{publisher}{Springer},
  \bibinfo{address}{Berlin}, \bibinfo{year}{1983}).

\bibitem[{\citenamefont{R.~Kubo}(1962)}]{kubo62}
\bibinfo{author}{\bibfnamefont{i.}~\bibnamefont{R.~Kubo}},
  \emph{\bibinfo{title}{{Fluctuations, Relaxation and Resonance in Magnetic
  Systems}}} (\bibinfo{publisher}{edited by D. ter Haar, Oliver and Boye},
  \bibinfo{address}{Edinburgh}, \bibinfo{year}{1962}).

\bibitem[{\citenamefont{{T. Yamaguchi}}(2000)}]{Yamaguchi}
\bibinfo{author}{\bibnamefont{{T. Yamaguchi}}}, \bibinfo{journal}{J. Chem.
  Phys.} \textbf{\bibinfo{volume}{112}}, \bibinfo{pages}{8530}
  (\bibinfo{year}{2000}).
  
\bibitem[{\citenamefont{{E. Gershgoren, Z. Wang, S. Ruhman, J.
  Vala, and R. Kosloff}}(2003)}]{k183}
\bibinfo{author}{\bibnamefont{{E. Gershgoren, Z. Wang, S. Ruhman,
  J. Vala, and R. Kosloff}}}, \bibinfo{journal}{J. Chem. Phys.}
  \textbf{\bibinfo{volume}{118}}, \bibinfo{pages}{3660} (\bibinfo{year}{2003}).
  
\bibitem[{\citenamefont{{M. Demirplak, S. Rice}}(2006)}]{Rice06}
\bibinfo{author}{\bibnamefont{{M. Demirplak, S. Rice}}}, \bibinfo{journal}{J.
  Chem. Phys.} \textbf{\bibinfo{volume}{125}}, \bibinfo{pages}{194517}
  (\bibinfo{year}{2006}).

\bibitem[{\citenamefont{{A. V. Uskov, A. P. Jauho, B. Tromborg, J. Mork and R.
  Lang}}(2000)}]{Uskov}
\bibinfo{author}{\bibnamefont{{A. V. Uskov, A. P. Jauho, B. Tromborg, J. Mork
  and R. Lang}}}, \bibinfo{journal}{Phys. Rev. Lett.}
  \textbf{\bibinfo{volume}{85}}, \bibinfo{pages}{1516} (\bibinfo{year}{2000}).

\bibitem[{\citenamefont{{C. Kammerer, C. Voisin, G. Cassabois, C. Delalande,
  Ph. Roussignol, F. Klopf, J. P. Reithmaier, A. Forchel and J. M.
  Gerard}}(2002)}]{Kammerer}
\bibinfo{author}{\bibnamefont{{C. Kammerer, C. Voisin, G. Cassabois, C.
  Delalande, Ph. Roussignol, F. Klopf, J. P. Reithmaier, A. Forchel and J. M.
  Gerard}}}, \bibinfo{journal}{Phys. Rev. B} \textbf{\bibinfo{volume}{66}},
  \bibinfo{pages}{041306(R)} (\bibinfo{year}{2002}).

\bibitem[{\citenamefont{{P. San-Jose, , G. Zarand, A. Shnirman and G.
  Schon}}(2002)}]{San-Jose}
\bibinfo{author}{\bibnamefont{{P. San-Jose, , G. Zarand, A. Shnirman and G.
  Schon}}}, \bibinfo{journal}{Phys. Rev. Lett.} \textbf{\bibinfo{volume}{97}},
  \bibinfo{pages}{076803} (\bibinfo{year}{2002}).

\bibitem[{\citenamefont{Iachello and Levine}(1995)}]{Iachello}
\bibinfo{author}{\bibfnamefont{F.}~\bibnamefont{Iachello}} \bibnamefont{and}
  \bibinfo{author}{\bibfnamefont{R. D.}~\bibnamefont{Levine}},
  \emph{\bibinfo{title}{{Algebraic Theory of Molecules}}}
  (\bibinfo{publisher}{Oxford University Press}, \bibinfo{address}{Oxford},
  \bibinfo{year}{1995}).

\bibitem[{\citenamefont{Perina}(1991)}]{Perina}
\bibinfo{author}{\bibfnamefont{J.}~\bibnamefont{Perina}},
  \emph{\bibinfo{title}{{Quantum Statistics of Linear and Nonlinear Optical
  Phenomena}}} (\bibinfo{publisher}{Kluwer}, \bibinfo{address}{London},
  \bibinfo{year}{1991}).

\bibitem[{\citenamefont{{D. J. Wineland, C. Monroe, W. M. Itano, D. Leibfried,
  B. E. King and D. M. Meekhof}}(1998)}]{Wineland98}
\bibinfo{author}{\bibnamefont{{D. J. Wineland, C. Monroe, W. M. Itano, D.
  Leibfried, B. E. King and D. M. Meekhof}}}, \bibinfo{journal}{J. Res. Nat.
  Inst. Stand. Tech.} \textbf{\bibinfo{volume}{103}}, \bibinfo{pages}{259}
  (\bibinfo{year}{1998}).

\bibitem[{\citenamefont{{E. Schmidt}}(1906)}]{schmidt}
\bibinfo{author}{\bibnamefont{{E. Schmidt}}}, \bibinfo{journal}{Math. Ann.}
  \textbf{\bibinfo{volume}{63}}, \bibinfo{pages}{433} (\bibinfo{year}{1906}).

\bibitem[{\citenamefont{Diosi}(2007)}]{Diosi07}
\bibinfo{author}{\bibfnamefont{L.}~\bibnamefont{Diosi}},
  \emph{\bibinfo{title}{{A Short Course in Quantum Information Theory}}}
  (\bibinfo{publisher}{Springer}, \bibinfo{address}{Berlin},
  \bibinfo{year}{2007}).

\bibitem[{\citenamefont{{M. Plenio, S. Virmani}}(2007)}]{virmani}
\bibinfo{author}{\bibnamefont{{M. Plenio, S. Virmani}}},
  \bibinfo{journal}{Quant. Inf. Comp.} \textbf{\bibinfo{volume}{7}},
  \bibinfo{pages}{1} (\bibinfo{year}{2007}).

\bibitem[{\citenamefont{{G. Vidal, R. F. Werner}}(2002)}]{vidal}
\bibinfo{author}{\bibnamefont{{G. Vidal, R. F. Werner}}},
  \bibinfo{journal}{Phys. Rev. A} \textbf{\bibinfo{volume}{65}},
  \bibinfo{pages}{032314} (\bibinfo{year}{2002}).

\bibitem[{\citenamefont{{A. Peres}}(1996)}]{peres96}
\bibinfo{author}{\bibnamefont{{A. Peres}}}, \bibinfo{journal}{Phys. Rev. Lett.}
  \textbf{\bibinfo{volume}{77}}, \bibinfo{pages}{1413} (\bibinfo{year}{1996}).

\bibitem[{\citenamefont{{M. Horodecki, P.Horodecki and
  R.Horodecki}}(1996)}]{Horodeckii}
\bibinfo{author}{\bibnamefont{{M. Horodecki, P.Horodecki and R.Horodecki}}},
  \bibinfo{journal}{Physics Letters A} \textbf{\bibinfo{volume}{223}},
  \bibinfo{pages}{1} (\bibinfo{year}{1996}).

\bibitem[{\citenamefont{{E. M. Rains }}(1999)}]{Rains}
\bibinfo{author}{\bibnamefont{{E. M. Rains }}}, \bibinfo{journal}{Phys. Rev. A}
  \textbf{\bibinfo{volume}{60}}, \bibinfo{pages}{179} (\bibinfo{year}{1999}).

\bibitem[{\citenamefont{{E.M. Rains }}(2001)}]{Rains_error}
\bibinfo{author}{\bibnamefont{{E.M. Rains }}}, \bibinfo{journal}{Phys. Rev. A}
  \textbf{\bibinfo{volume}{63}}, \bibinfo{pages}{019902(E)}
  (\bibinfo{year}{2001}).

\bibitem[{\citenamefont{{S. Virmani, M. F. Sacchi, M. B. Plenio and D. Markham
  }}(2001)}]{Virmani_sacchi}
\bibinfo{author}{\bibnamefont{{S. Virmani, M. F. Sacchi, M. B. Plenio and D.
  Markham }}}, \bibinfo{journal}{Physics Letters A}
  \textbf{\bibinfo{volume}{62}}, \bibinfo{pages}{288} (\bibinfo{year}{2001}).

\bibitem[{\citenamefont{{M. Khasin, R. Kosloff and D.
  Steinitz}}(accepted)}]{Dani}
\bibinfo{author}{\bibnamefont{{M. Khasin, R. Ronnie and D. Steinitz}}},
  \bibinfo{journal}{Phys. Rev. A} \textbf{\bibinfo{volume}{75}},
   (\bibinfo{year}{2007}). (\bibinfo{year}{accepted}).

\bibitem[{\citenamefont{Zurek}(1981)}]{Zurek81}
\bibinfo{author}{\bibfnamefont{W.~H.} \bibnamefont{Zurek}},
  \bibinfo{journal}{Phys. Rev. D} \textbf{\bibinfo{volume}{24}},
  \bibinfo{pages}{1516} (\bibinfo{year}{1981}).

\bibitem[{\citenamefont{Diosi and Kiefer}(2000)}]{lajos00}
\bibinfo{author}{\bibfnamefont{L.}~\bibnamefont{Diosi}} \bibnamefont{and}
  \bibinfo{author}{\bibfnamefont{C.}~\bibnamefont{Kiefer}},
  \bibinfo{journal}{Phys. Rev. Lett.} \textbf{\bibinfo{volume}{85}},
  \bibinfo{pages}{3552} (\bibinfo{year}{2000}).

\bibitem[{\citenamefont{Strunz}(2002)}]{strunz}
\bibinfo{author}{\bibfnamefont{W.~T.} \bibnamefont{Strunz}},
  \emph{\bibinfo{title}{Decoherence in Quantum Physics}} in \emph{\bibinfo{title}{Coherent Evolution in Noisy Environments}}
  (\bibinfo{publisher}{Springer, Berlin, Heidelberg}, \bibinfo{address}{edited
  by A. Buchleitner and K. Hornberger}, \bibinfo{year}{2002}).
  
\bibitem[{\citenamefont{Horn and Johnson}(1990)}]{Horn}
\bibinfo{author}{\bibfnamefont{R.}~\bibnamefont{Horn}} \bibnamefont{and}
  \bibinfo{author}{\bibfnamefont{C.~R.} \bibnamefont{Johnson}},
  \emph{\bibinfo{title}{{Matrix Analysis}}} (\bibinfo{publisher}{Cambridge
  University Press}, \bibinfo{address}{Cambridge}, \bibinfo{year}{1990}).

\bibitem[{\citenamefont{Gorini and Kossakowski}(1976)}]{gorini76}
\bibinfo{author}{\bibfnamefont{V.}~\bibnamefont{Gorini}} \bibnamefont{and}
  \bibinfo{author}{\bibfnamefont{A.}~\bibnamefont{Kossakowski}},
  \bibinfo{journal}{J. Math. Phys.} \textbf{\bibinfo{volume}{17}},
  \bibinfo{pages}{1298} (\bibinfo{year}{1976}).

\bibitem[{\citenamefont{{Emily A. Weiss, Gil Katz G, Randell H. Goldsmith,
  Michael R. Wasielewski, Mark A Ratner, Ronnie Kosloff,Abraham
  Nitzan}}(2006)}]{k216}
\bibinfo{author}{\bibnamefont{{E. A. Weiss, G. Katz , R. H.
  Goldsmith, M. R. Wasielewski, M. A. Ratner, R. Kosloff, A.
  Nitzan}}}, \bibinfo{journal}{J. Chem. Phys.} \textbf{\bibinfo{volume}{124}},
  \bibinfo{pages}{074501} (\bibinfo{year}{2006}).

\bibitem[{\citenamefont{Lindblad}(1976)}]{Lindblad76}
\bibinfo{author}{\bibfnamefont{G.}~\bibnamefont{Lindblad}},
  \bibinfo{journal}{Comm. Math. Phys.} \textbf{\bibinfo{volume}{48}},
  \bibinfo{pages}{119} (\bibinfo{year}{1976}).

\bibitem[{\citenamefont{{G. S. Agarwal and J. Banerji}}(1997)}]{Agarwal}
\bibinfo{author}{\bibnamefont{{G. S. Agarwal and J. Banerji}}},
  \bibinfo{journal}{Phys. Rev. A} \textbf{\bibinfo{volume}{55}},
  \bibinfo{pages}{R4007} (\bibinfo{year}{1997}).

\bibitem[{\citenamefont{{D. J. Wineland, C. Monroe, W. M. Itano, B. E. King, D.
  Leibfried, C. Myatt and C. Wood}}(1998)}]{Monroe98}
\bibinfo{author}{\bibnamefont{{D. J. Wineland, C. Monroe, W. M. Itano, B. E.
  King, D. Leibfried, C. Myatt and C. Wood}}}, \bibinfo{journal}{Physica
  Scripta} \textbf{\bibinfo{volume}{T76}}, \bibinfo{pages}{147}
  (\bibinfo{year}{1998}).

\bibitem[{\citenamefont{{M. Paternostro, M. S. Kim and P. L.
  Knight}}(2005)}]{Knight}
\bibinfo{author}{\bibnamefont{{M. Paternostro, M. S. Kim and P. L. Knight}}},
  \bibinfo{journal}{Phys. Rev. A} \textbf{\bibinfo{volume}{71}},
  \bibinfo{pages}{022311} (\bibinfo{year}{2005}).

\bibitem[{\citenamefont{{R. Grobe, K. Rzazewski, and J. H.
  Eberly}}(1994)}]{Eberly}
\bibinfo{author}{\bibnamefont{{R. Grobe, K. Rzazewski, and J. H. Eberly}}},
  \bibinfo{journal}{J. Phys B: At. Mol. Phys.} \textbf{\bibinfo{volume}{27}},
  \bibinfo{pages}{L503} (\bibinfo{year}{1994}).

\end{thebibliography}
   %\end{document}

\end{document}